\documentclass[aps,twocolumn,twoside,superscriptaddress]{revtex4}

\usepackage{currfile}
\usepackage[T1]{fontenc}                

\usepackage{graphicx,epsfig,verbatim,enumerate}
\usepackage{amssymb,amsmath}
\usepackage{url,hyperref}
\usepackage{doi}

\hypersetup{colorlinks=true,urlcolor=blue,citecolor=black,linkcolor=black}

\usepackage{natbib}

\usepackage{ifthen}
\newboolean{twocolswitch}

\newcommand{\sindex}[1]{}
\newcommand{\nindex}[1]{}

\newcommand{\www}[1]{\url{#1}}

\usepackage{lettrine}

\usepackage{color}

\newcommand{\havgfn}{h_{\rm avg}}

\newcommand{\revtexonly}[1]{#1}
\newcommand{\plainlatexonly}[1]{}

\setboolean{twocolswitch}{true}

\begin{document}

\title{
The Geography of Happiness:
Connecting
Twitter sentiment and expression,
demographics,
and 
objective characteristics of place
\\
\href{http://dx.plos.org/10.1371/journal.pone.0064417}{PLoS ONE, 8(5): e64417, 2013}
}

\author{
\firstname{Lewis}
\surname{Mitchell}
}

\email{lewis.mitchell@uvm.edu}
 \thanks{Corresponding author}


\author{
\firstname{Morgan R.}
\surname{Frank}
}

\email{morgan.frank@uvm.edu}


\author{
\firstname{Kameron Decker}
\surname{Harris}
}
  
\altaffiliation{
 Current address:
 Department of Applied Mathematics,
 University of Washington,
 Seattle, WA 98103 USA
}
\email{kamdh@uw.edu}


\author{
\firstname{Peter Sheridan}
\surname{Dodds}
}

\email{peter.dodds@uvm.edu}


\author{
\firstname{Christopher M.}
\surname{Danforth}
}
\email{chris.danforth@uvm.edu}

\affiliation{Computational Story Lab,
  Department of Mathematics \& Statistics,
  Vermont Complex Systems Center,
  \& the Vermont Advanced Computing Core,
  The University of Vermont,
  Burlington, 
  VT 05401.}

\date{May 29, 2013}

\begin{abstract}
We conduct a detailed investigation
of correlations between real-time expressions of 
individuals made across the United States and a wide range of 
emotional, geographic, demographic, and health characteristics.
We do so by combining (1) 
a massive, geo-tagged data set 
comprising  over 80 million words 
generated
in 2011
on the social network service Twitter 
and 
(2)
annually-surveyed characteristics
of all 50 states and close to 400 urban populations.
Among many results, 
we generate taxonomies of 
states and cities based 
on their similarities in word use;
estimate the happiness levels of states and cities;
correlate highly-resolved demographic characteristics
with happiness levels;
and
connect word choice and message length 
with urban characteristics such as education levels and obesity rates.
Our results show how social media may potentially be used
to estimate real-time levels and changes in population-level
measures such as obesity rates.
Extensive appendices of supplementary information for this work are maintained online at
\url{http://www.uvm.edu/storylab/share/papers/mitchell2013a}.
\end{abstract}

\maketitle


\section{Introduction}
\label{sec:geotweets.intro}

With vast quantities of real-time, fine-grained data, 
describing everything from transportation dynamics and
resource usage to
social interactions,
the science of cities has entered the realm of the data-rich fields.
While much work and development lies ahead, opportunities for quantitative
study of urban phenomena are now far
more broadly available to researchers~\cite{Bettencourt2007}.
With over half the world's population now living in urban areas, 
and this proportion continuing to grow, cities
will only become increasingly central to human society~\cite{Jacobs1961}.
Our focus here concerns one of the many important questions 
we are led to continuously address about cities:
how does living in urban areas relate to well-being?
Such an undertaking is part of a general program seeking to quantify 
and explain the evolving cultural character---the stories---of cities,
as well as geographic places of larger and smaller scales.



Numerous studies on well-being are published every year.
The UN's 2012 World Happiness
Report attempts to quantify happiness on a global scale using a `Gross
National Happiness' index which uses data on rural-urban residence and
other factors~\cite{Helliwell2012}.
In the US, Gallup and Healthways
produce a yearly report on the well-being of different cities,
states and congressional districts~\cite{GallupHealthways2012}, 
and they
maintain a well-being index based on continual 
polling and survey data~\cite{wellbeingindex_site}.
Other countries are also beginning to produce
measures of well-being: 
in 2012, surveys measuring national
well-being and how it relates to both health and where people live were
conducted in both the United Kingdom by the Office of National
Statistics~\cite{Thomas2012,Randall2012} and in Australia by Fairfax
Media and Lateral Economics~\cite{Lancy2013}.

While these and other approaches to quantifying the sentiment of a
city as a whole rely almost exclusively on survey data, 
there are now a range of 
complementary, remote-sensing methods available to researchers.
The explosion in the amount and
availability of data relating to social network use in the past 10
years has driven a rapid increase in the application of data-driven
techniques to the social sciences and sentiment analysis of
large-scale populations.

Our overall aim in this paper is to investigate how geographic place
correlates with and potentially influences societal levels of happiness.
In particular, after first examining happiness dynamics at the level of states,
we will explore urban areas in the United States in depth, 
and ask if it is possible to 
(a) 
measure the overall average happiness of people located in cities,
and
(b) 
explain the variation in happiness across different cities.
Our methodology for answering the first question uses word frequency
distributions collected from a large corpus of geolocated messages or
`tweets' posted on Twitter, with individual words scored for
their happiness independently by users of Amazon's Mechanical Turk
service~\cite{MechanicalTurk}.
This technique was introduced by Dodds
and Danforth (2009)~\cite{Dodds2009} and greatly expanded upon in
Dodds et al. (2011)~\cite{Dodds2011}, as well as tested for robustness
and sensitivity.
In attempting to answer the second question of 
happiness variability, we examine 
how individual word usage correlates with happiness
and various social and economic factors.
To do this we use the `word shift graph' technique
developed in~\cite{Dodds2009,Dodds2011}, as well as correlate word
usage frequencies with traditional city-level census survey data.
As we will show, the combination of these techniques 
produces significant insights into the character of 
different cities and places.


We structure our paper as follows.
In Section~\ref{sec:geotweets.dataandmethodology},
we describe the data sets and our methodology for measuring
happiness.
In Section~\ref{sec:geotweets.statescities}
we measure the happiness of different states and cities
and determine the happiest and saddest states and cities in the US, with some
analysis of why places vary with respect to this measure.
In Section~\ref{sec:geotweets.censusdata}
we compare our results for cities with census data, correlating happiness and
word usage with common social and economic measures.
We also use the
word frequency distributions to group cities by their similarities in
observed word use.
We conclude with a discussion in 
Section~\ref{sec:geotweets.discussion}.

\section{Data and methodology}
\label{sec:geotweets.dataandmethodology}

We examine a corpus of over 10 million geotagged tweets gathered from
373 urban areas in the contiguous United States during the calendar
year 2011.
This corpus is a subset of Twitter's `garden hose' feed, 
which in 2011 represented roughly 10\% of all messages. 
For the present study, 
we focus on the approximately 1\% of tweets that are geotagged.
Urban areas are defined by the 2010 
United States Census Bureau's MAF/TIGER
(Master Address File/Topologically Integrated Geographic Encoding and
Referencing) database~\cite{TIGER2010}.
Note that these urban area boundaries often agglomerate small towns together,
particularly when there are small towns geographically close to larger towns or cities.
See Appendix A for a more
detailed description of the data set as well as an exploration of the
relationship between area and perimeter, or fractal dimension, of
these cities.

To measure sentiment (hereafter happiness) in these areas from the
corpus of words collected, we use the Language Assessment by
Mechanical Turk (LabMT) word list (available online in the
supplementary material of~\cite{Dodds2011}), assembled by combining
the 5,000 most frequently occurring words in each of four text sources:
Google Books (English), music lyrics, the New York Times and
Twitter.
A total of roughly 10,000 of these individual words have been
scored by users of Amazon's Mechanical Turk service on a scale of 1
(sad) to 9 (happy), resulting in a measure of average happiness for
each given word~\cite{Kloumann2012}.
For example, `rainbow' is one of
the happiest words in the list with a score of $\havgfn = 8.1$, while
`earthquake' is one of the saddest, with $\havgfn = 1.9$.
Neutral words
like `the' or `thereof' tend to score in the middle of the scale, with
$\havgfn = 4.98$ and $5$ respectively.


For a given text $T$ containing $N$ unique words, we calculate the
average happiness $\havgfn$ by
\begin{equation}
  \havgfn(T) 
  =
  \frac{
    \sum_{i=1}^N \havgfn(w_i) f_i
  }
  {
    \sum_{i=1}^N f_i
  } 
  = 
  \sum_{i=1}^N \havgfn(w_i) p_i
  \label{eqn:havg}
\end{equation}
where $f_i$ is the frequency of the $i$th word $w_i$ in $T$ for which
we have a happiness value $\havgfn(w_i)$, and $p_i = f_i / \sum_{i=1}^N
f_i$ is the normalized frequency of word $w_i$.

Importantly, with this method we make no attempt to take the context
of words or the meaning of a text into account.
While this may lead to
difficulties in accurately determining the emotional content of small
texts, we find that for sufficiently large texts this approach
nonetheless gives reliable (if eventually improvable) results.
An analogy is that of temperature:
while the motion of a small number of particles cannot be expected to
accurately characterize the temperature of a room, an average over a
sufficiently large collection of such particles nonetheless defines a durable quantity.
Furthermore, by ignoring the context of words we gain
both a computational advantage and a degree of impartiality; we do not
need to decide \emph{a priori} whether a given word has emotional
content, thereby reducing the number of steps in the algorithm and
hopefully reducing experimental bias.

Following Dodds et al. (2011), for the remainder of this paper, we
remove all words $w_i$ for which the happiness score falls in the
range $4 < \havgfn(w_i) < 6$ when calculating $\havgfn(T)$.
Removal of
these neutral or `stop' words has been demonstrated to provide a
suitable balance between sensitivity and robustness in our
`hedonometer'~\cite{Dodds2011}.
Further details on how we preprocessed
the Twitter data set can be found in Appendix A.

We will correlate our happiness results with census data
which was taken from
the 2011 American Community Survey 1-year estimates, 
accessible online at \url{http://factfinder2.census.gov/}.

\section{Happiness across states and urban areas}
\label{sec:geotweets.statescities}

\begin{figure*}[tbp!]
  \begin{center}
    \includegraphics[width=2.0\columnwidth]{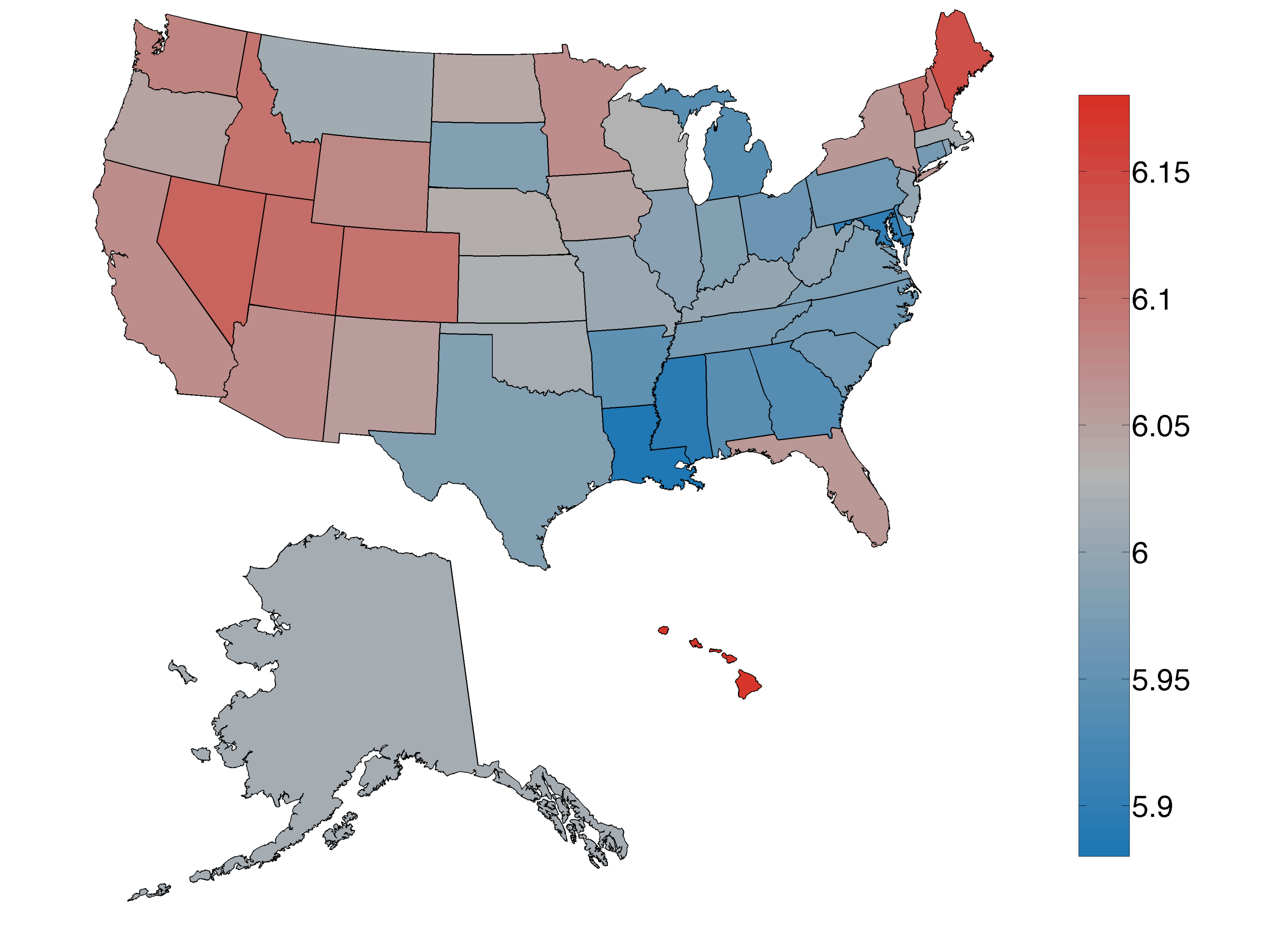}
\caption{
      Choropleth showing average word happiness for geotagged tweets
      in all US states collected during the calendar year 2011.  
      The
      happiest 5 states, in order, are: Hawaii,
      Maine,
      Nevada,
      Utah and Vermont.  
      The saddest 5 states, in order, are: Louisiana,
      Mississippi, Maryland, Delaware and Georgia.  
      Word shift plots
      describing how differences in word usage contribute to variation
      in happiness between states are presented in Appendix B
      (online)~\cite{geotweets-onlinematerial}.
          \label{fig:states}
    }
  \end{center}
\end{figure*}

We first examine how happiness varies on a somewhat coarser scale than
we will focus on for the majority of this paper, by plotting the
average happiness of all states in the US in Figure
\ref{fig:states}.
To avoid the problem that some states have happier
names than others, 
we removed each state name
from the calculation for $\havgfn$.
We also removed instances of the capitalized string `HI',
which generally occurred as the state code for Hawaii
and positively biased the score for that state.
We remark however that including this string
increased Hawaii's score by only 0.01; 
in general we find that the hedonometer is very robust 
to small variations in word frequencies such as this.

At such a
coarse resolution there is little variation between states, which all
lie between 0.15 of the mean value for the entire United States of
$\havgfn = 6.01$.
The happiest state is Hawaii with a score of $\havgfn =
6.16$ and the saddest state is Louisiana with a score of $\havgfn =
5.88$.
Hawaii emerges as the happiest state due to an abundance of
relatively happy words such as `beach' and food-related words. A similar result
showing greater happiness and a relative abundance of food-related
words in tweets made by users who regularly travel large distances (as
would be the case for many of the tweets emanating from Hawaii) has
been reported in~\cite{Frank2013}.
Louisiana is revealed as the
saddest state, with a significant factor being an abundance of profanity
relative to the other states.
This is in stark contrast with the findings
of Oswald and Wu~\cite{Oswald2010,Oswald2011},
who determined Louisiana to be the state with highest well-being
according to an alternate survey-based measure of life satisfaction.

\begin{figure*}[tbp!]
  \begin{center}
    \includegraphics[width=2.0\columnwidth]{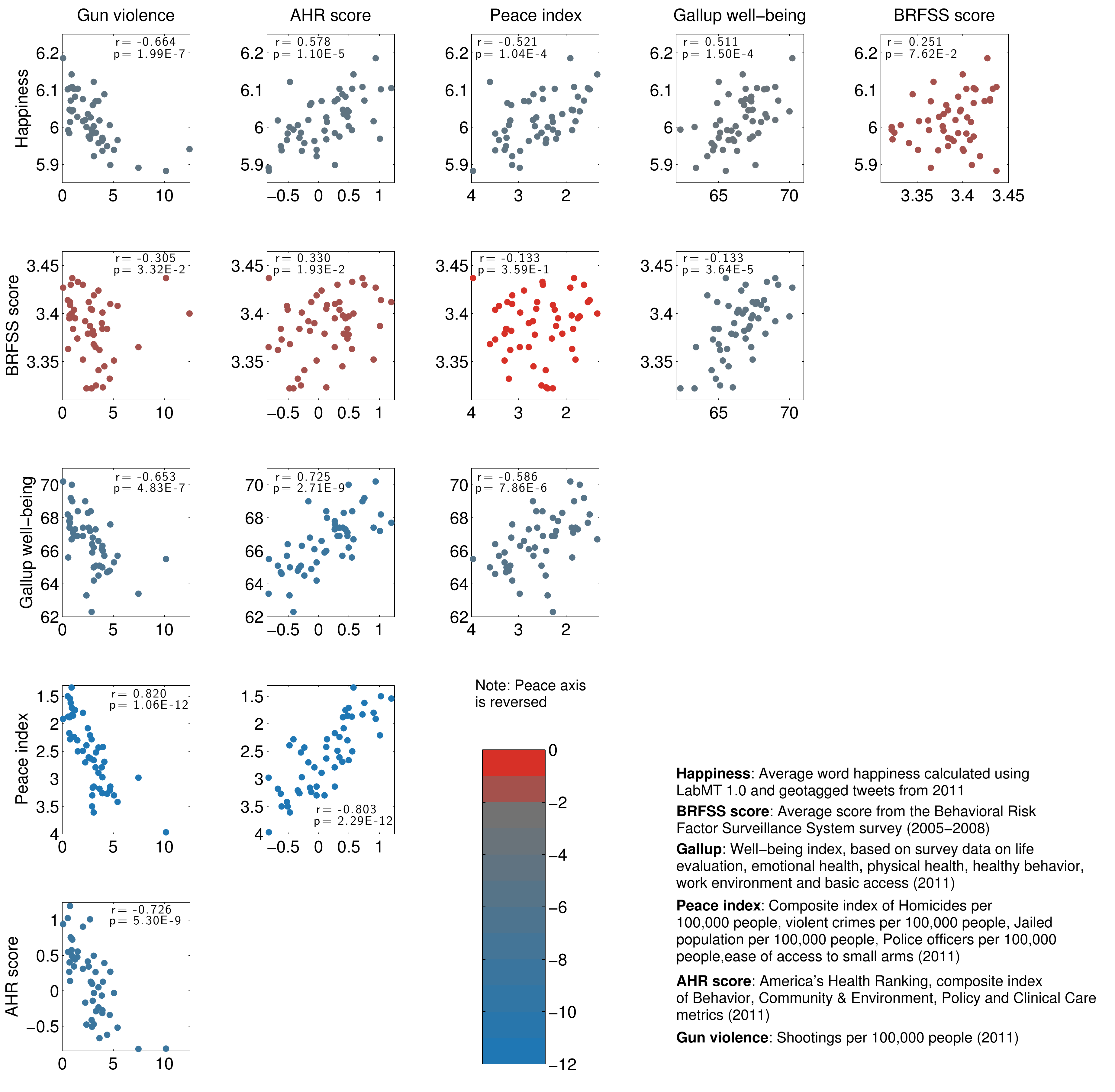}
    \caption{
    Scatter plot matrix of correlations between different well-being measures.
    Points are colored by $p$-value,
    statistically insignificant correlations above $p=0.01$ are shown in red.
    Spearman's $r$ and $p$-value are reported in the inset.
        \label{fig:happinessScatterMatrix}
    }
  \end{center}
\end{figure*}

In Figure~\ref{fig:happinessScatterMatrix} we compare our results with five other well-being measures:
\begin{itemize}
\item the behavioral risk factor survey score (BRFSS) used by Oswald and Wu~\cite{Oswald2011},
a survey of life satisfaction across the United States;
\item the 2011 Gallup well-being index~\cite{GallupHealthways2012}, 
based on survey data about life evaluation, emotional and physical health,
healthy behavior, work environment and basic access;
\item the 2011 United States peace index~\cite{USPI2011} 
produced by the Institute for Economics and Peace, 
a composite index of homicides per
100,000 people, violent crimes per 100,000 people, 
size of jailed population per 100,000 people, 
number of police officers per 100,000 people 
and ease of access to small arms;
\item the 2011 United Health Foundation's America's health ranking (AHR)~\cite{AHR2011}, 
a composite index of behavior, 
community and environment, 
policy and clinical care metrics;
\item the number of shootings per 100,000 people in 2011.
\end{itemize}

Figure \ref{fig:happinessScatterMatrix} shows a matrix of scatter plots 
showing the correlations between 
each of the above measures, 
including average word happiness.
Spearman's $r$ and $p$-values are reported in the inset for each 
scatter plot.
Points are colored by $p$-value,
with blue points indicating stronger correlation 
and red indicating insignificant correlations above $p=0.01$.
Our measure of state happiness (top row) correlates strongly 
with all other measures
except for the BRFSS,
however the BRFSS itself correlates significantly only with the Gallup well-being index.
Possible explanations for the poor agreement between BRFSS and the other measures
may include its placing of Louisiana at the top of the BRFSS well-being list,
which is generally opposite to its position in similar lists.
The BRFSS also uses data collected between 2005 and 2008,
whereas all the other lists use data from 2011 only.

\begin{figure*}[tbp!]
  \begin{center}
    \includegraphics[width=2.0\columnwidth]{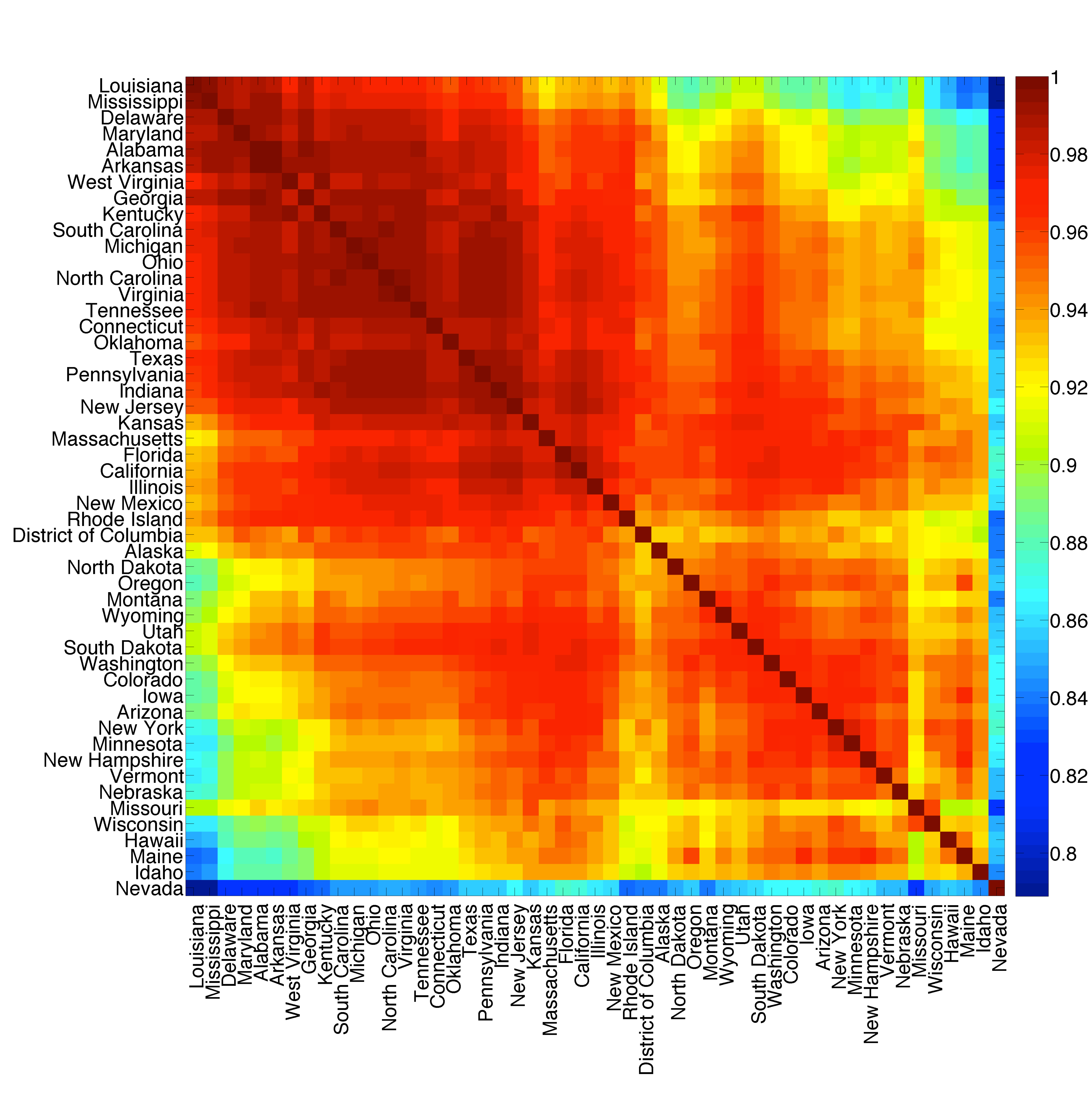}
    \caption{
      Clustergram showing cross-correlations between word frequency
      distributions for all states in 2011.
      Red signifies states with
      similar or highly-correlating word frequency distributions,
      while blue signifies states with relatively dissimilar word
      frequency distributions.
        \label{fig:statesclustergram}
    }
  \end{center}
\end{figure*}

\begin{figure*}[tbp!]
  \begin{center}
    \includegraphics[width=2.0\columnwidth]{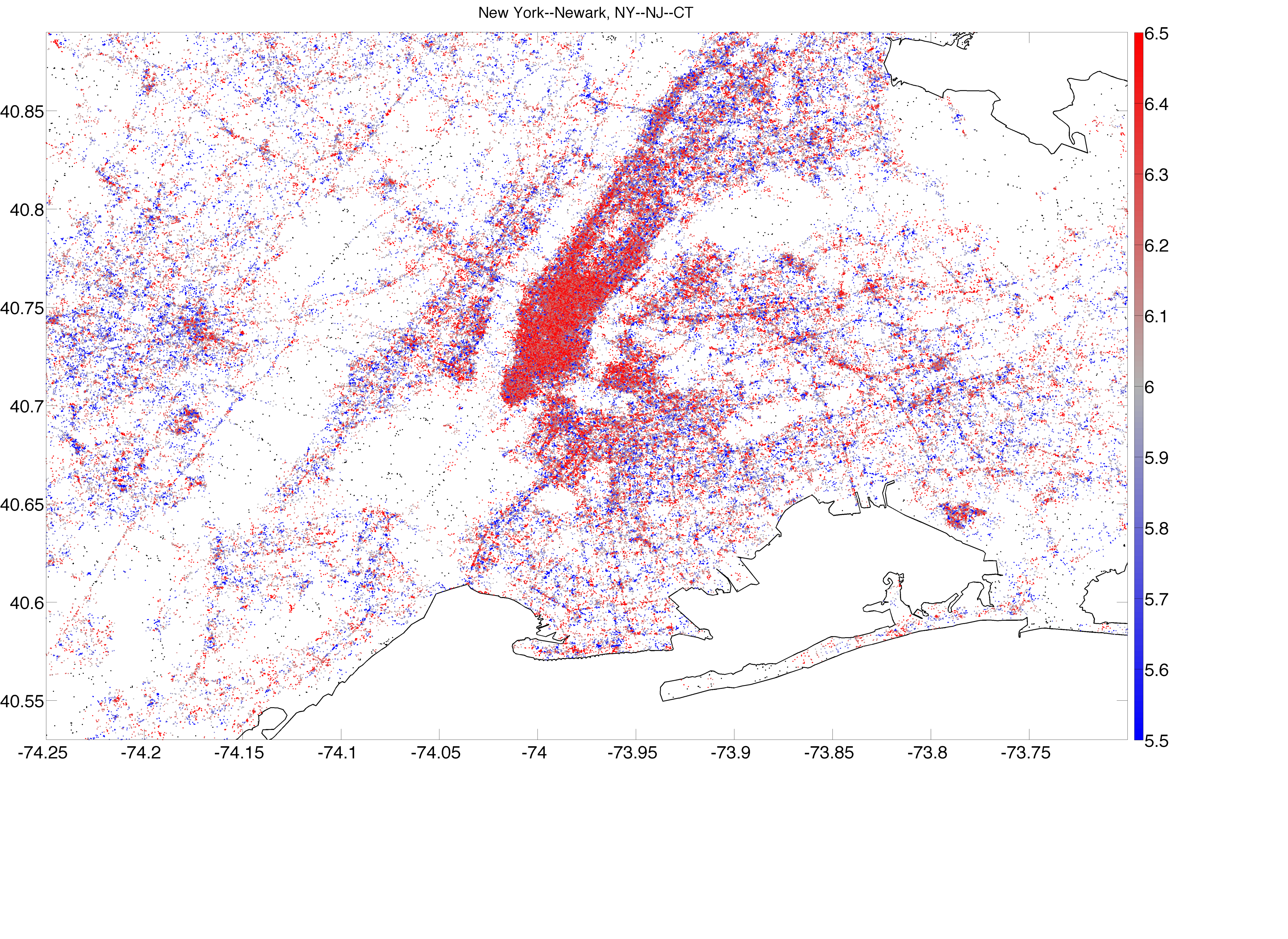}
    \caption{
      Map of tweets collected from New York City during the calendar
      year 2011.
      Each point represents an individual tweet and is
      colored by the average word happiness $\havgfn$ of nearby tweets:
      red is happier, blue is sadder.
      For a point to be colored, we
      require that there be at least 200 LabMT words within a 500
      meter radius of the location; points which do not satisfy this
      criterion are colored black. Maps for all other cities can be found in Appendix C (online)~\cite{geotweets-onlinematerial}.
          \label{fig:NYC}
    }
  \end{center}
\end{figure*}

We can further use this data on word frequencies to characterize
similarities between states based on word usage.
For simplicity, 
we focus on the 50,000 most frequently occurring words on Twitter~\cite{Dodds2011}.
Figure \ref{fig:statesclustergram} shows the
linear correlation between word frequency vectors ${\bf f} = \{f_i,
i=1:50000\}$ for each pair of states, with red entries in the matrix
indicating states with similar word use.
We see some clusters which
might be explained by geographical proximity, such as Vermont and New
Hampshire or Louisiana and Mississippi, and some outliers such as the
state of Nevada, which correlates the lowest on average with all other
states.
Additional details on this state-level dataset, including
plots of raw number of tweets and number of tweets per head of
population for each state can be found in Appendix A. 
Word shift graphs showing which words contribute most to the
variation in happiness across states can be found in Appendix B (online)~\cite{geotweets-onlinematerial}.

We now change our resolution to a finer scale by focussing on cities
rather than states.
As an illustration of the resolution of the data
set as well as our technique, we plot a tweet-generated map of a city,
showing how average word happiness varies with location.
In Figure
\ref{fig:NYC} we plot tweets collected from the New York City area
during 2011.
Each point represents an individual tweet, and is colored
by the happiness $\havgfn$ of the text $T$ consisting of the $N=200$
LabMT words contained in the geotagged tweets closest to that location.
We set a maximum
threshold radius of $r = 500$ meters within which to find other geotagged tweets
around each point;
if 200 LabMT words cannot be found 
within that radius then the point is
colored black.

Several features can immediately be discerned in this
purely tweet-generated map.
Firstly, the spatial resolution reveals
the outline of Manhattan, as well as Central Park, individual streets
and bridges, and even airport terminals such as those at JFK and
Newark airports at the lower right and center left of the figure
respectively.
Secondly, we can discern regions of higher and lower
happiness: the Harlem and Washington Heights areas to the north appear
relatively sad compared to the Downtown/Midtown area, as does the
Waterfront, New Jersey area west of the southern tip of Manhattan.
Similar tweet-generated maps for all 373 cities measured are
presented in Appendix B (online)~\cite{geotweets-onlinematerial}.

In Figure \ref{fig:USA_2011} we show a tweet-generated happiness map
of the entire contiguous United States, where we have now used $N =
500$ and $r = 10$ km.
We can clearly discern cities and the roads
between them at this scale, and substantial variation in happiness
across geographical regions.
There is already an indication that some
cities will be significantly less happy than others, particularly
those in the southeastern United States, a conclusion which will be
made more quantitative later.
At a finer scale we can see that some
coastal areas, particularly around the Florida peninsula and along the
coast of North and South Carolina, are significantly happier than the
regions immediately inland of them.
We will see this again below in
the word shifts for various oceanside cities.
Finally, we remark upon one limitation of the present methodology by noting 
that the Mexican cities shown in Figure \ref{fig:USA_2011}
appear far sadder than their counterparts to the north.
This is due to the presence of Spanish words such as `con' and `sin',
which while neutral in Spanish have been scored as negative
English words in LabMT.
At present the LabMT list is applicable only to English-language texts; 
future versions of the list will incorporate scores for languages other than English as well.

\revtexonly{\begin{turnpage}}
  \revtexonly{\begin{figure*}[tbp!]}
    \plainlatexonly{\begin{figure}[tbp!]}
      \centering
      \includegraphics[width=1.0\textheight]{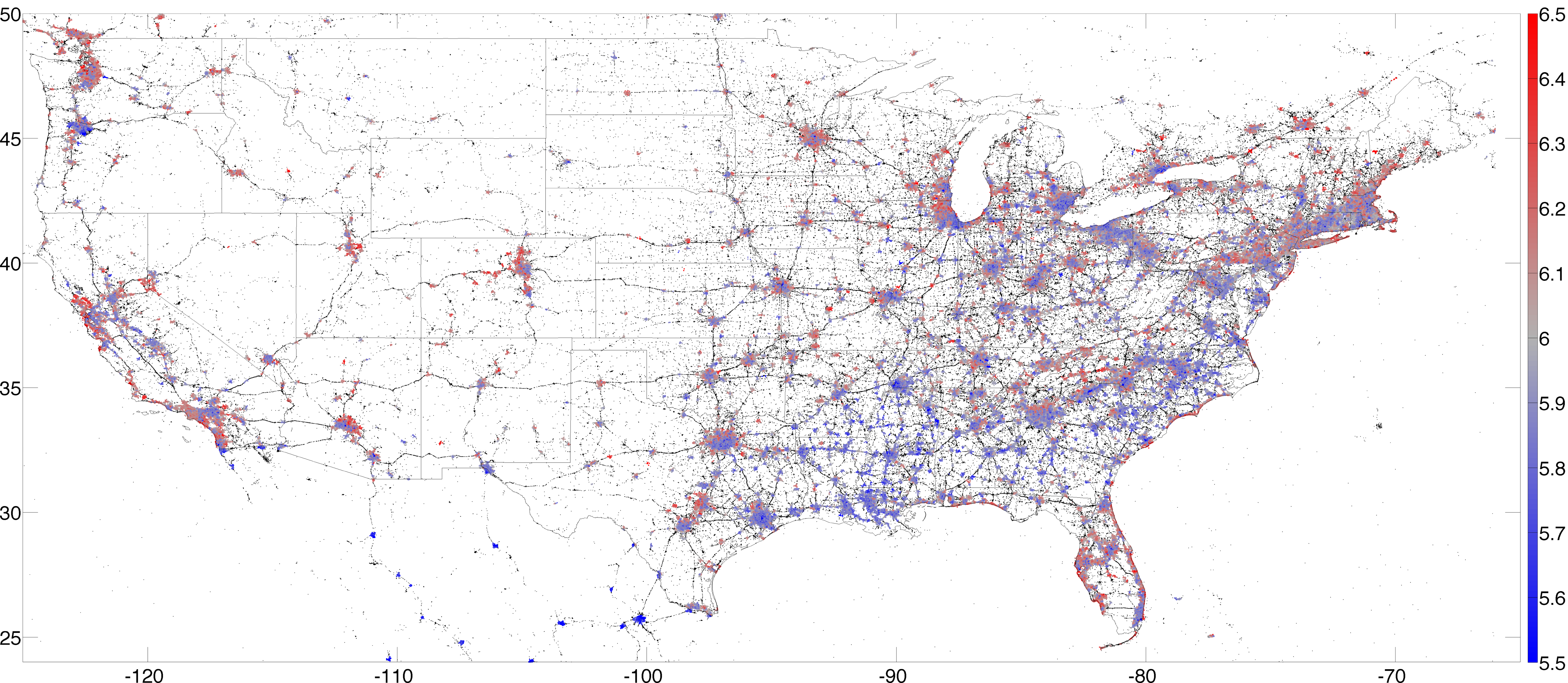}
      \caption{
        Map showing happiness of all tweets collected from the lower 48
        states during the calendar year 2011.
        Points are colored as in
        Figure \ref{fig:NYC}, except we now require that there are at
        least 500 LabMT words within a 10 kilometer radius of the
        location of each tweet in order to be colored. 
      \label{fig:USA_2011}
      }
    \plainlatexonly{\end{figure}}
  \revtexonly{\end{figure*}}
\revtexonly{\end{turnpage}}

\begin{figure}[tbp!]
  \begin{center}
    \includegraphics[width=\columnwidth]{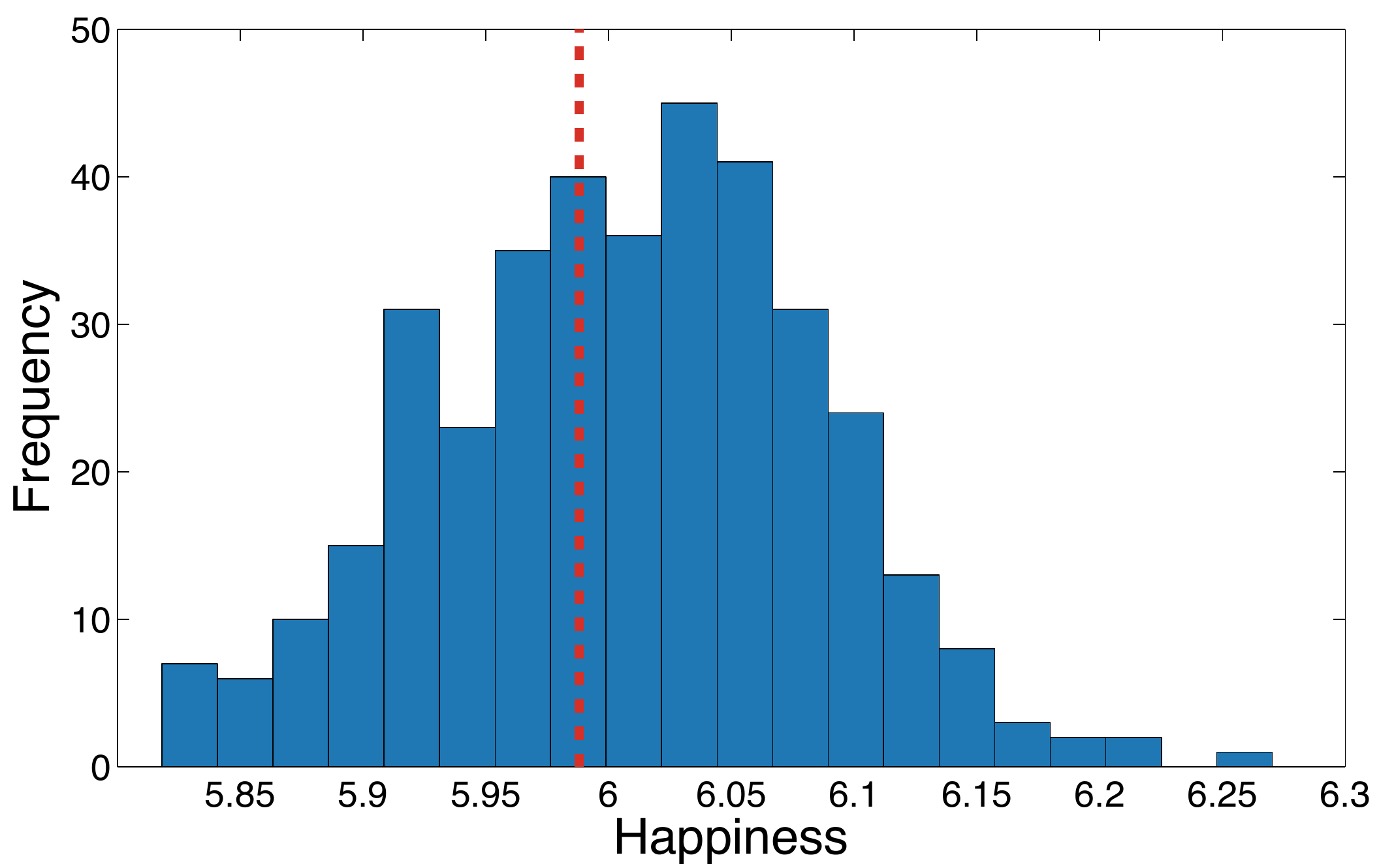}
    \caption{
      Histogram showing the distribution of happiness values for the
      373 cities in the census data set.
A vertical dashed line
      denotes the average for all cities.
Note the greater weight
      towards the right of the distribution, with more cities having
      happiness scores higher than the average.   
         \label{fig:city_hap_hist}
}
  \end{center}
\end{figure}

Next we calculate the happiness $\havgfn$ for each city in the census
data set using equation (\ref{eqn:havg}), where the boundaries of a
city are defined by the MAF/TIGER database, and each text $T$ is
formed by agglomerating all the words falling within that city.
Figure \ref{fig:city_hap_hist} shows the distribution of happiness
scores for all cities; as is to be expected for smaller samples, the
range of values is slightly higher than that calculated for the
states, extending over a range of more than 0.2 from the mean of
$\havgfn = 6.00$.
We remark that the distribution is skewed: there are
more cities that are happier than the overall average, by 220 to 153.

\begin{figure}[tbp!]
  \begin{center}
    \includegraphics[width=\columnwidth]{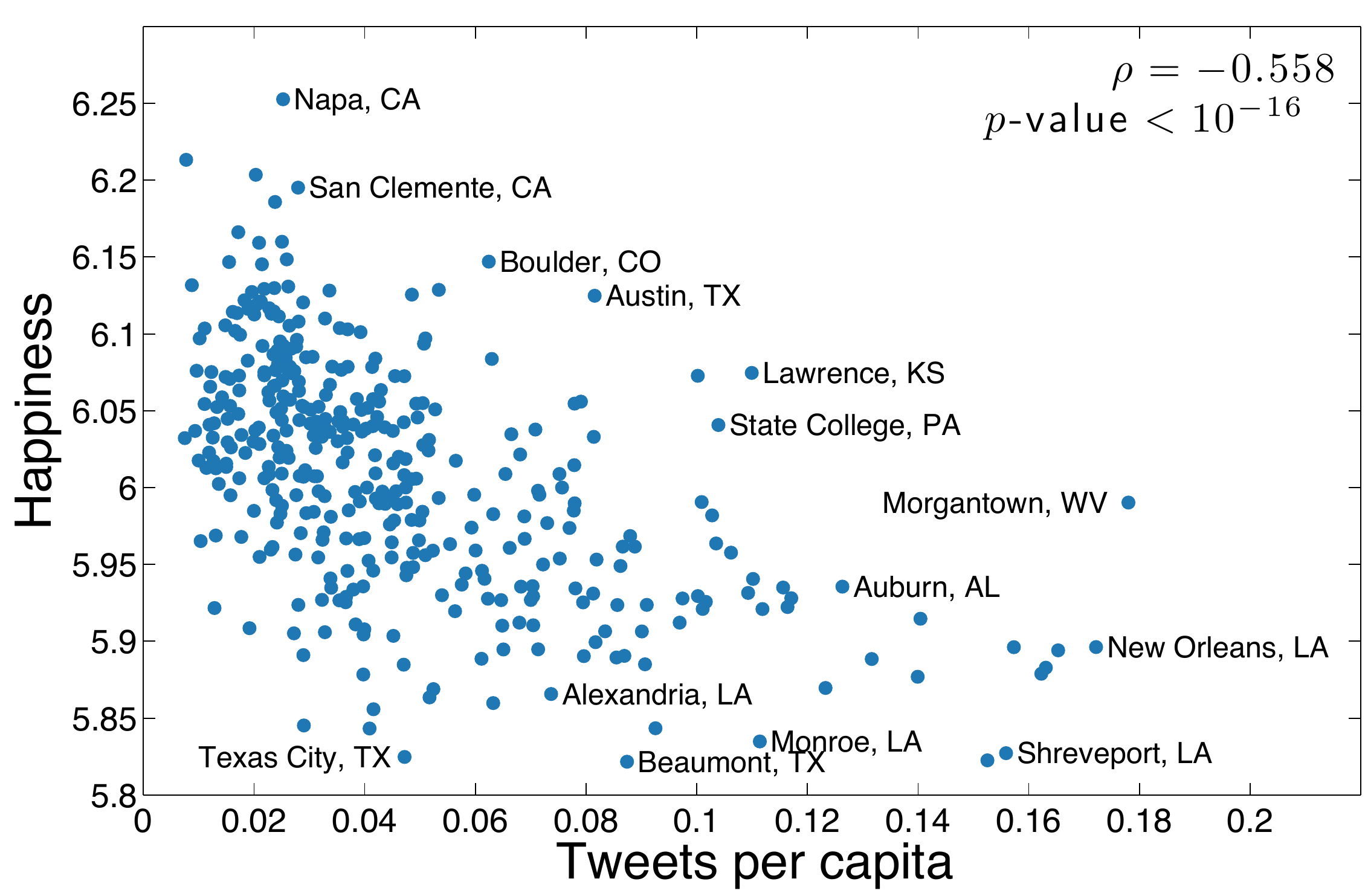}
    \caption{
      Happiness as a function of number of tweets per capita.
Areas
      with a higher density of tweets per capita tend to be less
      happy.  
          \label{fig:tweetspercapita_happiness}
          }
  \end{center}
\end{figure}

It is well known that city population sizes follow a power law
distribution (see~\cite{Zipf1949} and many others), which in
conjunction with Figure \ref{fig:city_hap_hist} suggests that
happiness decreases with city size.
While we do find a slight negative 
correlation between happiness and the number of tweets gathered in
each city, we in fact find that happiness more strongly negatively
correlates with the number of tweets per capita, with Spearman
correlation coefficient -0.558 and $p$-value less than $10^{-16}$, as
shown in Figure \ref{fig:tweetspercapita_happiness}.

\begin{figure}[tbp!]
  \begin{center}
    \includegraphics[width=\columnwidth]{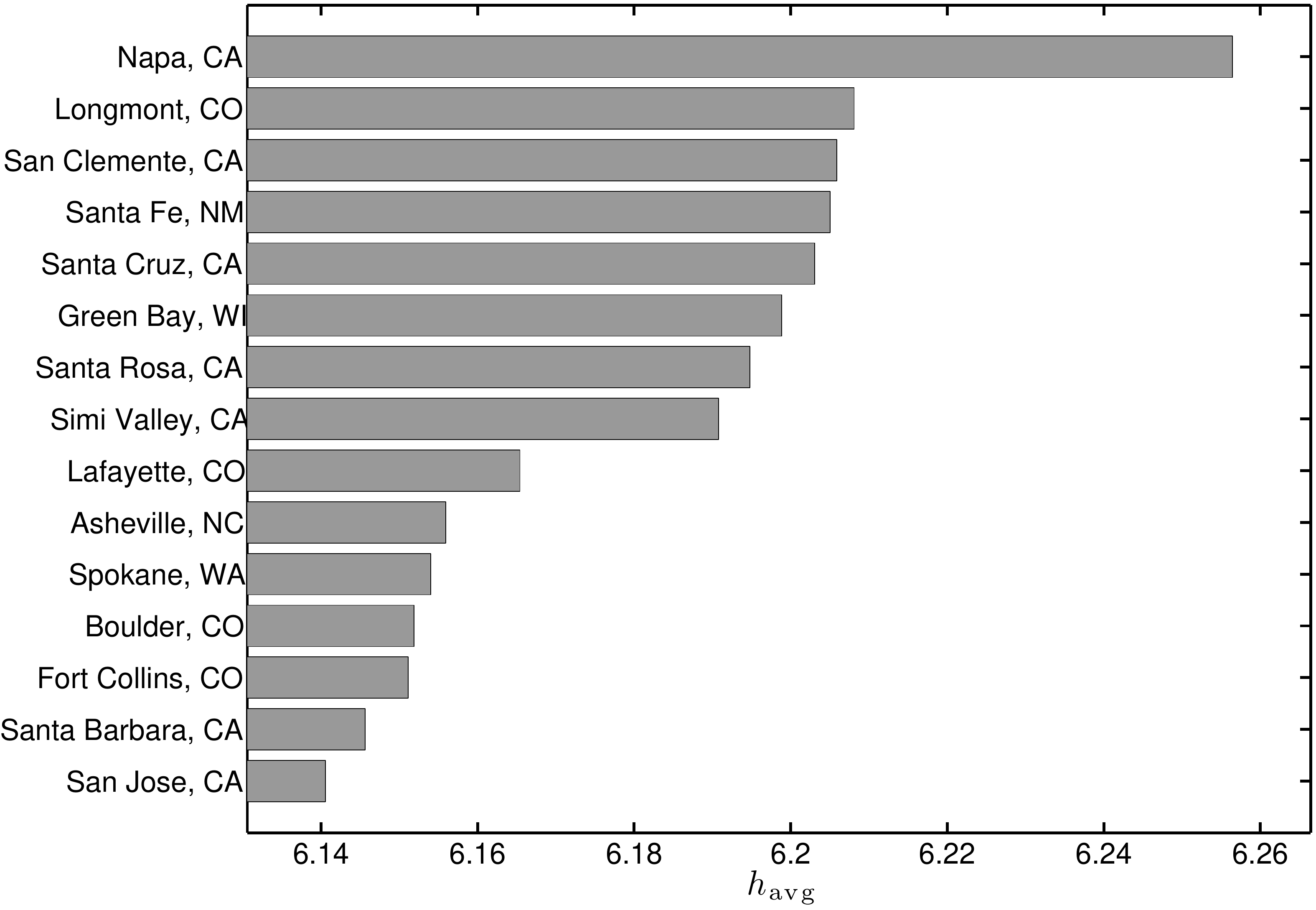}
    \caption{
      The 15 highest average word happiness scores $\havgfn$ for cities
      in the contiguous USA, as calculated using (\ref{eqn:havg}) and
      the LabMT word list. 
      The full list of cities can be found in Appendix C (online)~\cite{geotweets-onlinematerial}.
          \label{fig:happiest_15}
}
  \end{center}
\end{figure}

\begin{figure}[tbp!]
  \begin{center}
    \includegraphics[width=\columnwidth]{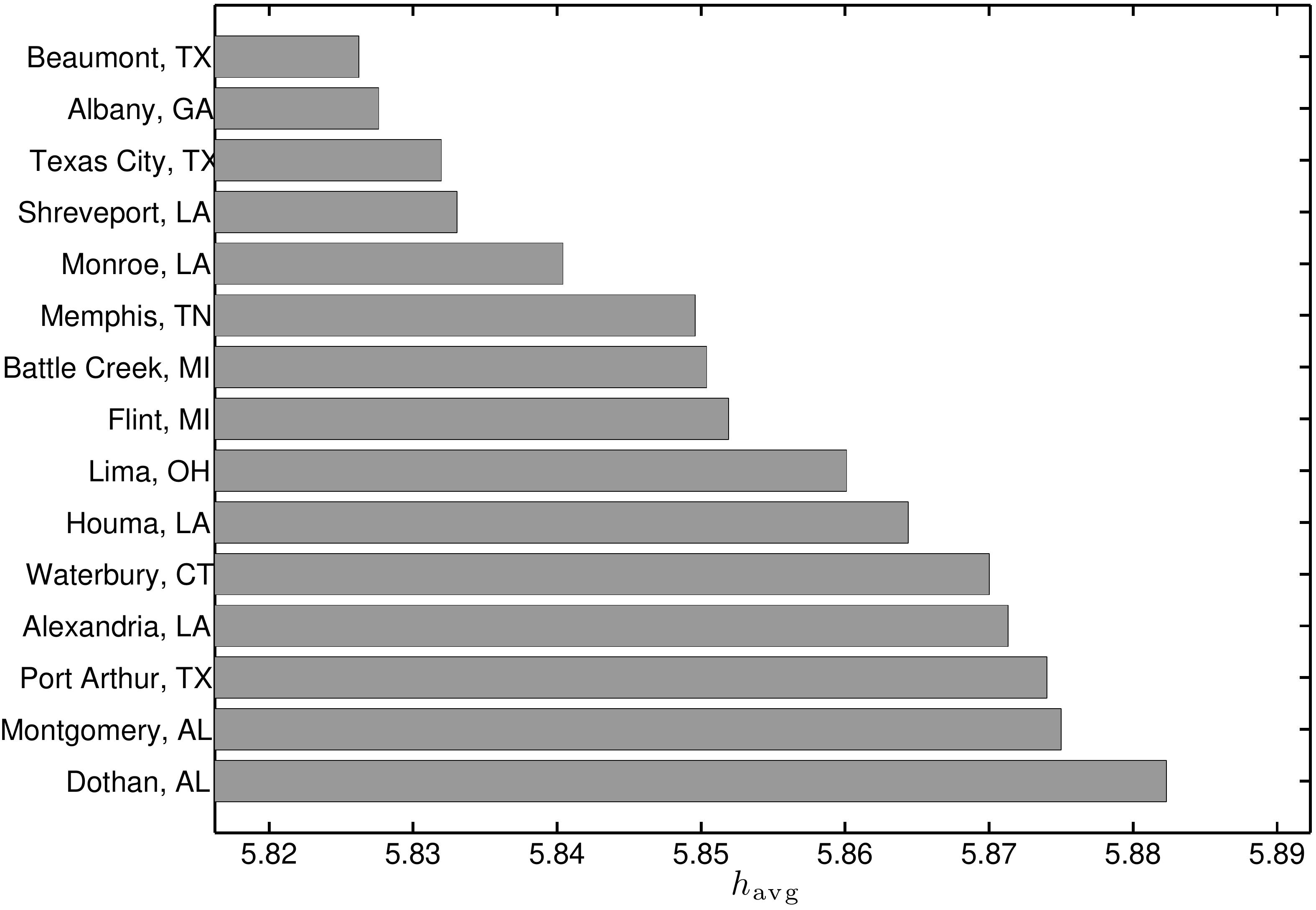}
    \caption{
      The 15 lowest average word happiness scores $\havgfn$ for cities
      in the contiguous USA, as calculated using (\ref{eqn:havg}) and
      the LabMT word list. 
            The full list of cities can be found in Appendix C (online)~\cite{geotweets-onlinematerial}.
        \label{fig:saddest_15}
         }
  \end{center}
\end{figure}

The bar charts in Figures \ref{fig:happiest_15} and
\ref{fig:saddest_15} show the average word happiness $\havgfn$ for the
15 happiest and 15 saddest cities in the contiguous United States,
respectively.
Using this method we identify Napa, California as the
happiest city in the US with a score of 6.26, and Beaumont, Texas as
the saddest city with a score of 5.83.

As was the case with our state happiness rankings, 
several cities that ranked both highly and lowly
by our measure rank similarly in more traditional survey based
efforts.
For example, the 2011 Gallup-Healthways well-being survey~\cite{GallupHealthways2012} showed Boulder, Colorado as the city with
the fifth highest well-being index composite score (and twelfth
highest happiness score in our list), while Flint, Michigan had the
second lowest and Montgomery, Alabama the 21st-lowest well-being index
(compared to 8th lowest and 14th lowest happiness scores on our list).
The overall Spearman correlation between the rankings using Gallup's
well-being index and our measure is $r = 0.328$, with
$p$-value $7.73 \times 10^{-6}$ (a scatter plot is presented online in
Appendix C).
Whereas our list uses only word frequencies in the
calculation of $\havgfn$, the Gallup-Healthways score is an average of
six indices which measure life evaluation, emotional health, work
environment, physical health, healthy behaviors, and access to basic
necessities.
We remark that our method is far more efficient to
implement than a survey-based approach, and it provides a near
real-time stream of information quantifying well-being in cities.

\begin{figure*}[tbp!]
  \begin{center}
    \includegraphics[width=\columnwidth]{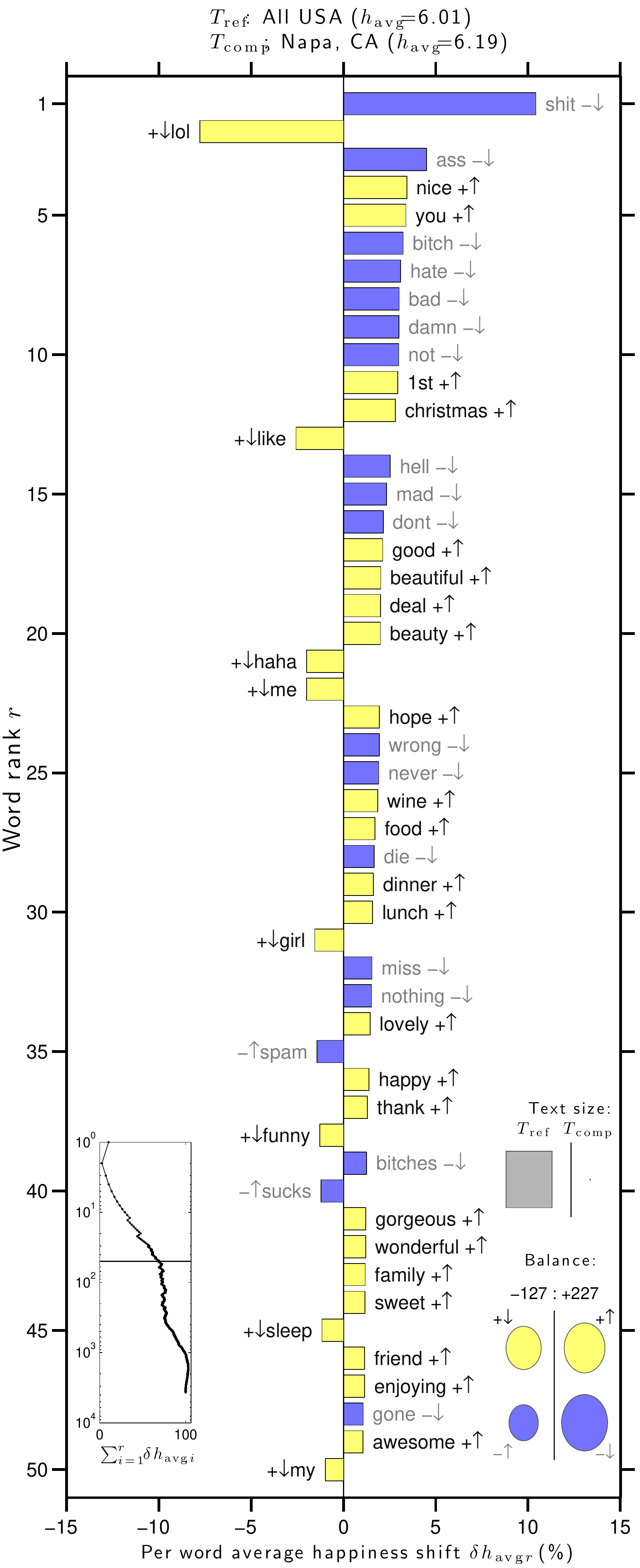}
    \includegraphics[width=\columnwidth]{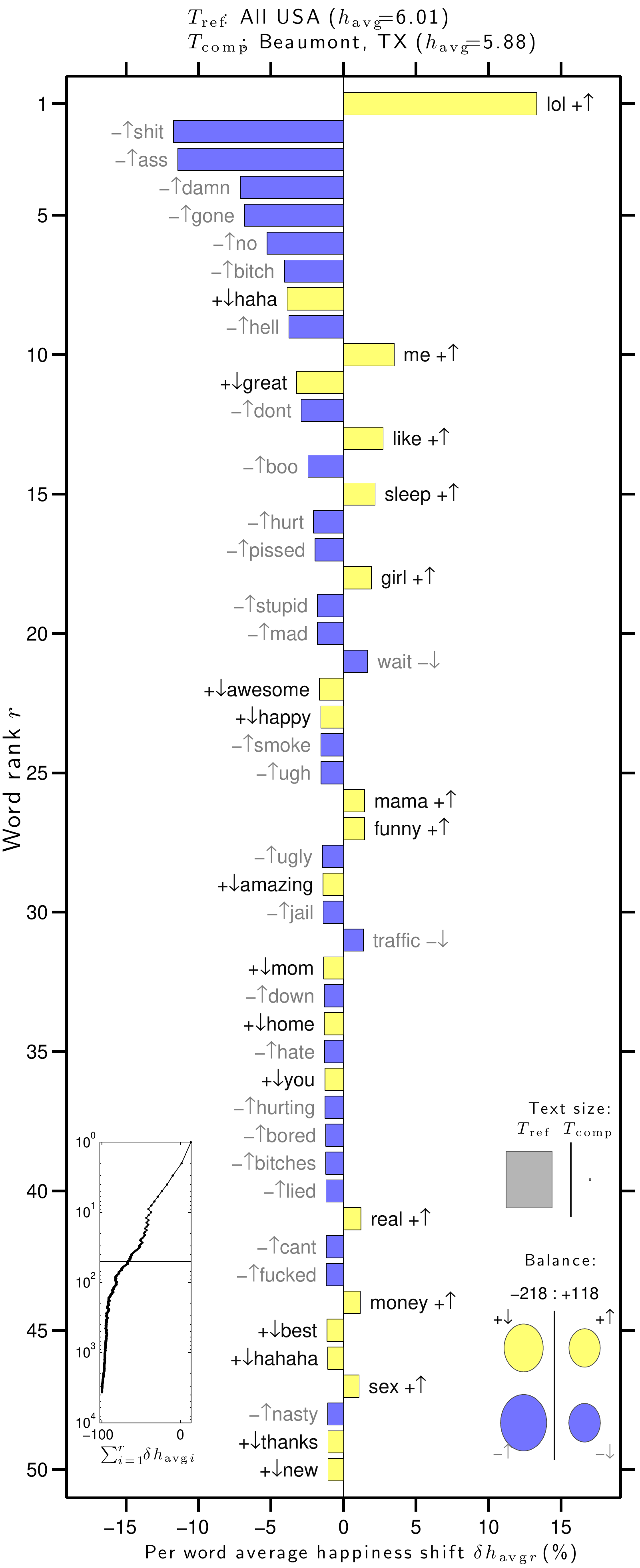}
    \caption{
      Word shift graphs showing how $\havgfn$ varies for all US cities
      measured versus the cities Napa, California (left) and Beaumont,
      Texas (right) with highest and lowest $\havgfn$ respectively.
      Words are ranked in order of decreasing percentage contribution
      to the overall average happiness difference $\delta \havgfn$.
The
      symbols $+/-$ indicate whether a word is relatively happy or sad
      compared to $\havgfn$ for the entire US (text $T_{\rm ref}$),
      while the arrows $\uparrow / \downarrow$ indicate whether the
      word was used more or less in the text $T_{\rm comp}$ for each
      city than in $T_{\rm ref}$.
The left inset panel shows how the
      ranked LabMT words combine in sum.
The four circles at bottom
      right show the total contribution of the four kinds of words
      ($+\downarrow$, $+\uparrow$, $-\uparrow$, $-\downarrow$).
      Relative text size is indicated by the areas of the gray
      squares.
          \label{fig:wordshifts}
}
  \end{center}
\end{figure*}

To investigate why the average word happiness varies across urban
areas, we study the word shift graphs~\cite{Dodds2009,Dodds2011} for
each city.
These graphs show how the difference in happiness for two
texts depends on differences in the underlying word frequencies.
In
Figure \ref{fig:wordshifts} we show the word shift graphs for Napa and
Beaumont, as compared to the entire corpus of words collected for all
urban areas during 2011.
Word shift graphs for every city are
presented in Appendix C (online)~\cite{geotweets-onlinematerial}.

We observe some features of the graphs that are consistent with
geography---for example the word `beach' appears high on the list of
words for coastal cities such as Santa Cruz, California or Miami,
Florida.
Overall, the main factor driving the relative happiness
scores for each city appears to be the presence or absence of key
words such as `lol', `haha' and its variants, `hell', `love', `like' 
and the negative words `no', `don't', `never' and `wrong',
as well as profanity.


\section{Correlating word usage with census data}
\label{sec:geotweets.censusdata}

The word shifts of Figure \ref{fig:wordshifts} demonstrate how word
usage varies with location, as well as the importance of studying the
individual words that go in to the calculation of averaged quantities
such as the word happiness $\havgfn$.
We therefore now examine in
greater detail how happiness and word usage relate to underlying
social factors.

We first focus on how the average happiness $\havgfn$ correlates with
different social and economic measures.
To do this we took data from
the 2011 American Community Survey 1-year estimates, specifically
tables DP02 through DP05 covering selected social characteristics,
economic characteristics, housing characteristics and demographic and
housing estimates.
These tables contained 508 different categories
for all cities, from which we removed the categories with data on less
than 75\% of all cities, leaving 432 different categories for
correlation with happiness.

\begin{figure*}[tbp!]
  \begin{center}
    \includegraphics[width=2.0\columnwidth]{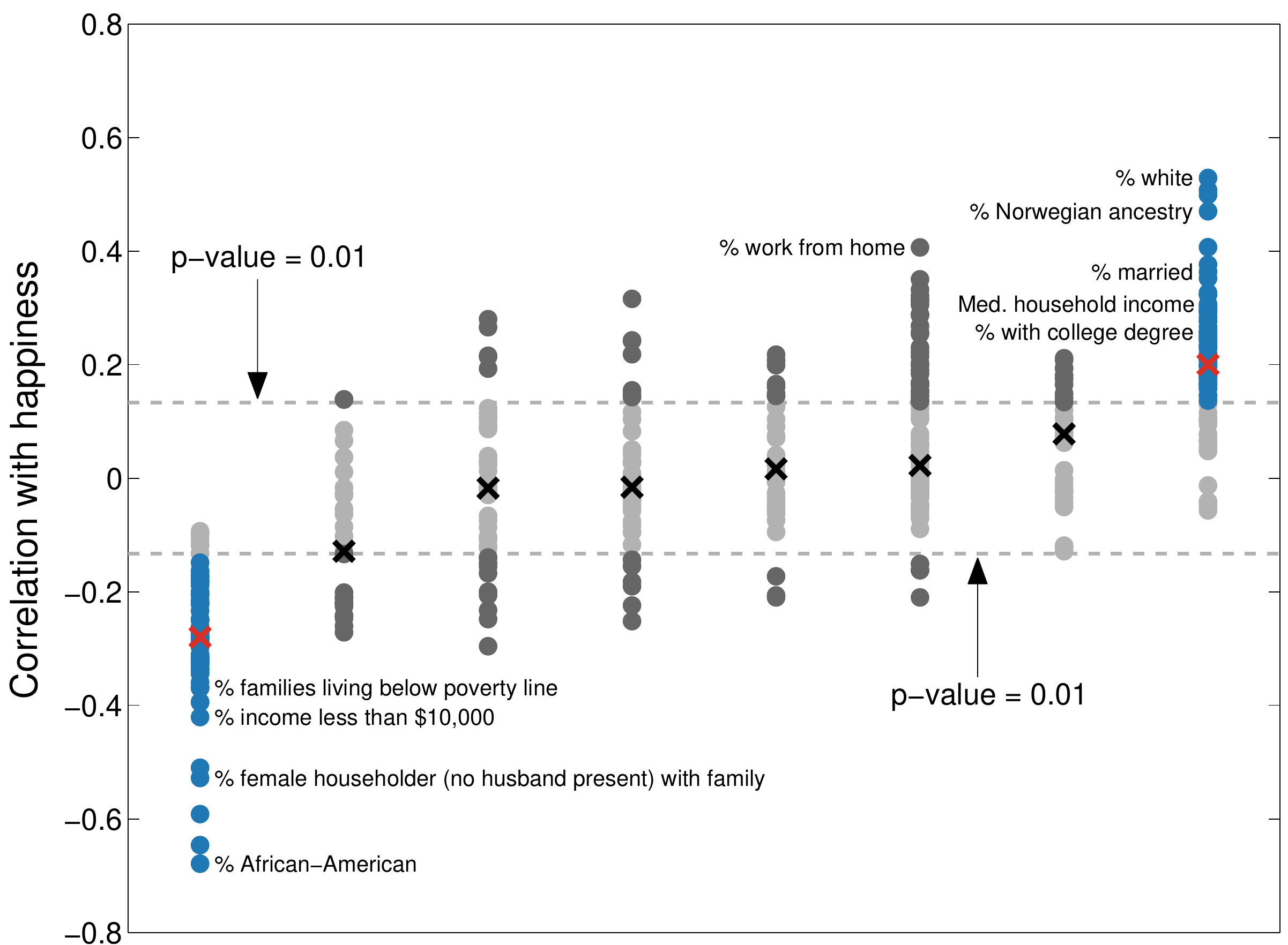}
    \caption{
      Spearman correlations for 432 demographic attributes with
      happiness.
The 8 groupings along the horizontal axis are for
      covarying attributes identified by agglomerative hierarchical
      clustering, independently of happiness.
Crosses lie on the
      median of each cluster, and the dashed lines represent the 1\%
      significance level.
The two clusters which have medians that
      correlate significantly with happiness are colored blue.
A
      complete list of the correlation of all attributes with
      happiness can be found in Appendix D (online)~\cite{geotweets-onlinematerial}..
          \label{fig:demographics_happiness}
}
  \end{center}
\end{figure*}

In Figure \ref{fig:demographics_happiness} we show the Spearman
correlation between happiness and each demographic attribute for
all 373 cities.
Each point in the graph represents one
of the 432 attributes considered; a table listing each demographic and
its correlation with happiness is presented in Appendix D (online)~\cite{geotweets-onlinematerial}.
The groupings into columns were made independently of happiness
values, by performing complete-link clustering using a hierarchical
cluster tree on the table of census attributes for all cities~\cite{Jain1999}.
The 8
clusters found are not unique and depend on the distance
threshold used, however they give some indication of which attributes
covary.
Only two groups show a large number of attributes which
significantly correlate (below $p=0.01$) with happiness; these are
shown in blue (with red crosses specifying the median attribute).
These two groups might be broadly characterized as representing high
socioeconomic and low socioeconomic status respectively, with many of
the attributes in the high socioeconomic status group positively
correlating with happiness,
and anti-correlating for the low socioeconomic
status group.

\begin{figure}[tbp!]
  \begin{center}
    \includegraphics[width=\columnwidth]{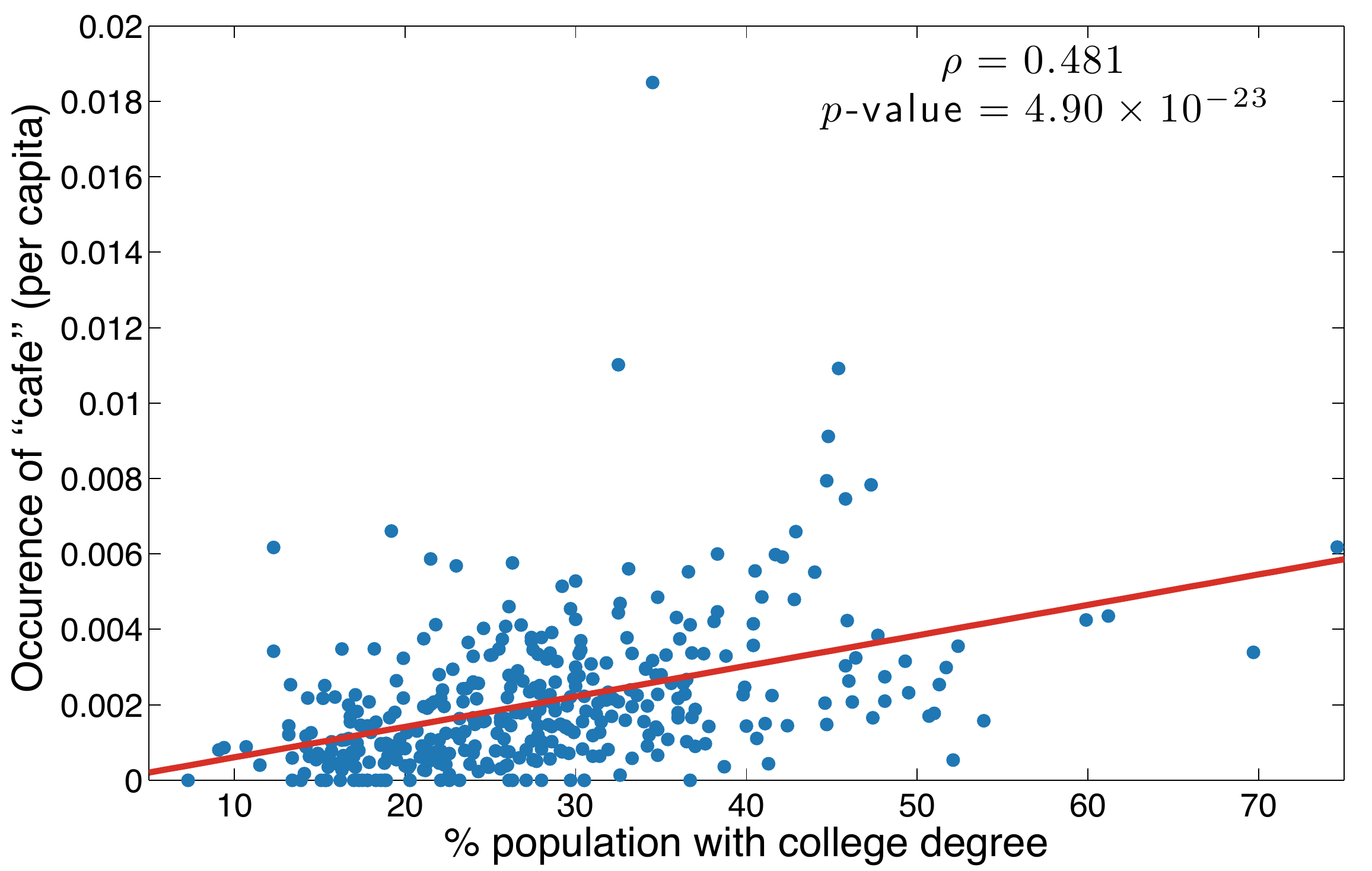}
    \caption{
      Scatter plot showing the correlation between rate of occurrence
      of the word `cafe' and percentage of population with a
      bachelor's degree or higher in US cities during the calendar
      year 2011.
The red line shows linear correlation while the
      reported $r$ and $p$-values show the Spearman correlation. 
          \label{fig:film_bachelors}
 }
  \end{center}
\end{figure}

To further understand what drives this correlation of certain
demographics with happiness, we now investigate how each word from the
LabMT list correlates with each census attribute.
To do this
we first normalize the word counts in each urban area by the total
number of tweets collected in each city, and then for each word
calculate the Spearman correlation $r$ between normalized frequency
and census attribute for all cities.
For example, the scatter plot in
Figure \ref{fig:film_bachelors} shows that the normalized frequency of
occurrence of the word `cafe' shows a strong positive correlation with
the percentage of the population with a bachelors degree or higher.
The Spearman correlation between the two is $r = 0.481$ with
$p$-value $4.90 \times 10^{-23}$, indicating strong correlation.

We present lists showing the correlation of each LabMT word with every
demographic attribute in Appendix D (online)~\cite{geotweets-onlinematerial}.
Taking
the percentage of population with a bachelors degree or higher for
urban areas from the 2011 census as a representative example, Tables
\ref{tbl:poscorrpovstopwords} and \ref{tbl:negcorrpovstopwords} show
the top 25 words which exhibit the highest positive and negative
correlations respectively with this attribute.
We note that the positive correlations in Table \ref{tbl:poscorrpovstopwords} are much
stronger than the negative correlations in Table \ref{tbl:negcorrpovstopwords};
a similar asymmetry appears in many of the tables in Appendix D.
The results show that
longer words such as `software', `development' and `emails' correlate
strongly with education, while the words which correlate negatively
with high levels of education are generally shorter, with no words longer than two
syllables appearing in the list.
Furthermore, many of the words such
as `love', `talk' and `mom' appearing in Table
\ref{tbl:negcorrpovstopwords} are family- or relationship-oriented,
while the words in Table
\ref{tbl:poscorrpovstopwords} are generally more employment-oriented, and
suggest more complex and abstract intellectual themes.
It may be
postulated that this is a reflection of the social processes occurring
in urban areas characterized by low and high education rates,
respectively.

\begin{table}[htb]
  \begin{center}
    \begin{tabular}{|l|c|c|c|}
      \hline Word & $r$ & $p$-value & $\havgfn(w_i)$\\ \hline cafe &
      0.481 & $4.9 \times 10^{-23}$ & 6.78\\ pub & 0.463 & $3.14
      \times 10^{-21}$ & 6.02\\ software & 0.458 & $9.07 \times
      10^{-21}$ & 6.30\\ yoga & 0.455 & $1.85 \times 10^{-20}$ &
      7.04\\ grill & 0.433 & $1.78 \times 10^{-18}$ & 6.24\\
      development & 0.424 & $1.14 \times 10^{-17}$ & 6.38\\ emails &
      0.419 & $2.87 \times 10^{-17}$ & 6.54\\ wine & 0.417 & $3.83
      \times 10^{-17}$ & 6.42\\ library & 0.414 & $6.47 \times
      10^{-17}$ & 6.48\\ art & 0.414 & $6.8 \times 10^{-17}$ & 6.60\\
      sciences & 0.410 & $1.54 \times 10^{-16}$ & 6.30\\ pasta & 0.410
      & $1.57 \times 10^{-16}$ & 6.86\\ lounge & 0.409 & $1.68 \times
      10^{-16}$ & 6.50\\ market & 0.408 & $2.2 \times 10^{-16}$ &
      6.28\\ india & 0.407 & $2.5 \times 10^{-16}$ & 6.42\\ drinking &
      0.405 & $3.74 \times 10^{-16}$ & 6.14\\ technology & 0.405 &
      $3.76 \times 10^{-16}$ & 6.74\\ forest & 0.405 & $3.83 \times
      10^{-16}$ & 6.68\\ brunch & 0.405 & $3.89 \times 10^{-16}$ &
      6.32\\ dining & 0.403 & $4.92 \times 10^{-16}$ & 6.48\\
      supporting & 0.399 & $1.1 \times 10^{-15}$ & 6.48\\ professor &
      0.398 & $1.23 \times 10^{-15}$ & 6.04\\ university & 0.392 &
      $3.62 \times 10^{-15}$ & 6.74\\ film & 0.391 & $4.27 \times
      10^{-15}$ & 6.56\\ global & 0.391 & $4.72 \times 10^{-15}$ &
      6.00\\ \hline
    \end{tabular}
  \end{center}
  \caption{
    Top 25 words with strongest positive Spearman correlation $r$
    to percentage of population with a Bachelors degree or higher
    (census table DP02-HC03-VC94) in 2011.
Stop words with $4 < \havgfn
    < 6$ have been removed from the list.
Note the low $p$-values for
    all words, indicating strong statistical significance. 
      \label{tbl:poscorrpovstopwords}
 }
\end{table}

\begin{table}[htb]
  \begin{center}
    \begin{tabular}{|l|c|c|c|}
      \hline Word & $r$ & $p$-value & $\havgfn(w_i)$\\ \hline me &
      -0.393 & $3.26 \times 10^{-15}$ & 6.58\\ love & -0.389 & $6.51
      \times 10^{-15}$ & 8.42\\ my & -0.354 & $1.97 \times 10^{-12}$ &
      6.16\\ like & -0.346 & $6.04 \times 10^{-12}$ & 7.22\\ hate &
      -0.344 & $8.76 \times 10^{-12}$ & 2.34\\ tired & -0.343 & $1
      \times 10^{-11}$ & 3.34\\ sleep & -0.341 & $1.27 \times
      10^{-11}$ & 7.16\\ stupid & -0.328 & $8.55 \times 10^{-11}$ &
      2.68\\ bored & -0.315 & $5.11 \times 10^{-10}$ & 3.04\\ you &
      -0.315 & $5.23 \times 10^{-10}$ & 6.24\\ goodnight & -0.305 &
      $1.77 \times 10^{-9}$ & 6.58\\ bitch & -0.295 & $6.51 \times
      10^{-9}$ & 3.14\\ all & -0.289 & $1.33 \times 10^{-8}$ & 6.22\\
      lie & -0.285 & $2.24 \times 10^{-8}$ & 2.60\\ mom & -0.284 &
      $2.42 \times 10^{-8}$ & 7.64\\ wish & -0.271 & $1.05 \times
      10^{-7}$ & 6.92\\ talk & -0.267 & $1.74 \times 10^{-7}$ & 6.06\\
      she & -0.265 & $2.01 \times 10^{-7}$ & 6.18\\ know & -0.262 &
      $2.78 \times 10^{-7}$ & 6.10\\ ill & -0.259 & $4.11 \times
      10^{-7}$ & 2.42\\ dont & -0.258 & $4.54 \times 10^{-7}$ & 3.70\\
      well & -0.256 & $5.3 \times 10^{-7}$ & 6.68\\ don't & -0.255 &
      $5.8 \times 10^{-7}$ & 3.70\\ give & -0.255 & $5.84 \times
      10^{-7}$ & 6.54\\ friend & -0.255 & $6.27 \times 10^{-7}$ &
      7.66\\ \hline
    \end{tabular}
  \end{center}
  \caption{
    Top 25 words with strongest negative Spearman correlation $r$
    to percentage of population with a Bachelors degree or higher in
    2011 (with stop words removed).  
      \label{tbl:negcorrpovstopwords}
}
\end{table}

The technique applied here is not limited only to census data.
As an example of a
different use of the corpus, we now correlate word use to obesity
at the metropolitan level.
For this study we take obesity levels from
the Gallup and Healthways 2011 survey~\cite{Witters2012}, and
metropolitan areas as defined by the U.S.
Office of Management and
Budget's Metropolitan Statistical Areas (MSAs)
\cite{MSAs}.
These MSAs are generally two to three times
larger in area than the TIGER urban area census boundaries, and the
Gallup obesity survey was only for the 190 largest-population areas.
The obesity data set therefore contains fewer small cities than the TIGER census
set does, particularly in the Midwest.
We collected more than 10 million
tweets from these 190 MSAs, corresponding to just over 80 million
words during 2011.

\begin{figure}[tbp!]
  \begin{center}
    \includegraphics[width=\columnwidth]{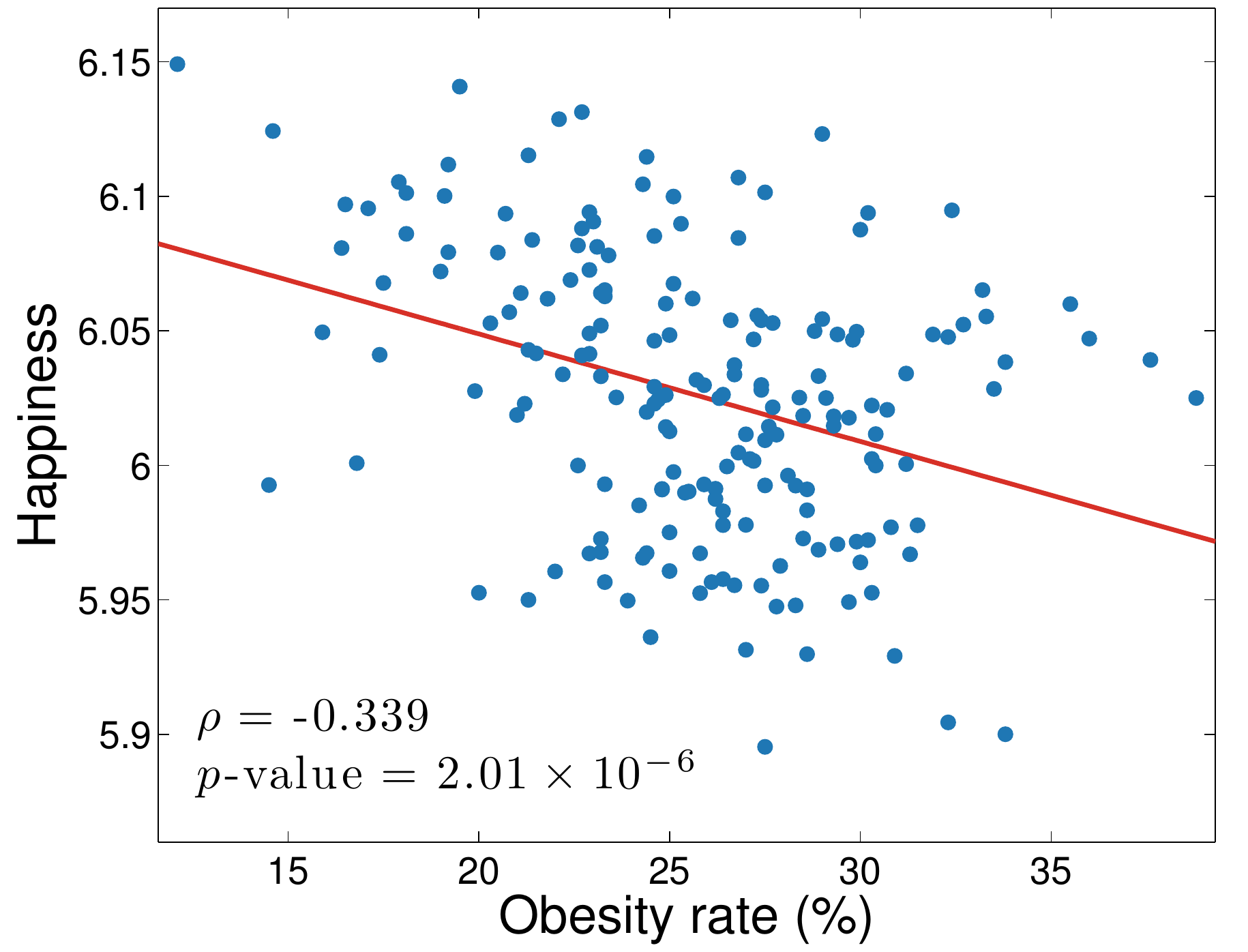}
    \caption{
      Scatter plot showing correlation between $\havgfn$ and obesity
      level, as taken from the 2011 Gallup and Healthways survey.  
      The red line is the straight line of best fit to the data, while the
      $r$ value is the Spearman correlation coefficient for the
      data.  
          \label{fig:obesity_happiness}
}
  \end{center}
\end{figure}

Performing the same analysis as for the attributes in Figure
\ref{fig:demographics_happiness}, in Figure
\ref{fig:obesity_happiness} we show the relationship between happiness
and obesity for the 190 MSAs included in the Gallup survey.
We find
that happiness generally decreases as obesity increases, with the
third happiest city in this set (Boulder, Colorado) corresponding with the
lowest obesity rate (12.1\%) and the saddest city (Beaumont, Texas, as
found previously) corresponding with the fifth highest obesity rate
(33.8\%).
We calculate a Spearman correlation coefficient ($r =
-0.339$ with $p$-value $2.01\times 10^{-6}$) 
which indicates statistically significant negative correlation
between obesity and happiness.

As we did for the census data, we also correlate the abundance of
each individual word in the LabMT list to obesity levels in the 190
cities surveyed.
From this list we extract words that are clearly
food-related, and in Table
\ref{tbl:obesitynegpos} present those which most most strongly correlate
(both negatively
and positively) with obesity.
Note that we are including stop words for
which $ 4 < \havgfn(w_i) < 6$ in these lists.
Coffee-related words such
as `cafe', `coffee', `espresso' and `bean' feature prominently in the
list, and many of the words refer to eating at restaurants---`sushi',
`restaurant', `cuisine' and `brunch', for example.
As we might expect
such words to correlate with wealth, this suggests a correlation
between obesity and poverty, a claim which we note remains contentious
in the medical literature (for example, supported in
\cite{Levine2011,Hruschka2012a}, and refuted in~\cite{Chang2005}).

Conversely, only 6 food-related words significantly positively
correlate with obesity with $p$-values less than 0.05 (note again the
asymmetry in the number of words which positively and negatively
correlate with obesity).
The fast food chain `mcdonalds' correlates
most strongly, and the foods `wings' and `ham' both appear.
Unlike in
the low-obesity word table, words describing a desire for food---`eat'
and `hungry'---as well as the negative reaction of `heartburn' to
overeating, both appear on the list.
In Appendix A we show tables
listing the food-related words which show the least correlation with
obesity, as well as the top 25 words (food-related or not) from the
LabMT list that correlate and anti-correlate with obesity. 
The full list of LabMT words and their correlations with obesity can be found in 
Appendix E (online)~\cite{geotweets-onlinematerial}.

\begin{table}[h]
  \begin{center}
    \begin{tabular}{|l|c|c|c|}
      \hline Word & $r$ & $p$-value & $\havgfn(w_i)$\\ \hline cafe &
      -0.509 & $6.07 \times 10^{-14}$ & 6.78\\ sushi & -0.487 & $9.93
      \times 10^{-13}$ & 5.40\\ brewery & -0.469 & $8.67 \times
      10^{-12}$ & N/A\\ restaurant & -0.448 & $8.93 \times 10^{-11}$ &
      7.06\\ bar & -0.435 & $3.59 \times 10^{-10}$ & 5.82\\ banana &
      -0.434 & $3.77 \times 10^{-10}$ & 6.86\\ apple & -0.408 & $5.22
      \times 10^{-9}$ & 7.44\\ fondue & -0.403 & $8.34 \times 10^{-9}$
      & N/A\\ wine & -0.400 & $1.08 \times 10^{-8}$ & 6.42\\ delicious
      & -0.392 & $2.17 \times 10^{-8}$ & 7.92\\ dinner & -0.386 &
      $3.85 \times 10^{-8}$ & 7.40\\ coffee & -0.384 & $4.51 \times
      10^{-8}$ & 7.18\\ bakery & -0.383 & $5.12 \times 10^{-8}$ &
      N/A\\ bean & -0.378 & $7.88 \times 10^{-8}$ & 5.80\\ espresso &
      -0.377 & $8.47 \times 10^{-8}$ & N/A\\ cuisine & -0.376 & $8.82
      \times 10^{-8}$ & N/A\\ foods & -0.374 & $1.07 \times 10^{-7}$ &
      7.26\\ tofu & -0.372 & $1.27 \times 10^{-7}$ & N/A\\ brunch &
      -0.368 & $1.79 \times 10^{-7}$ & 6.32\\ veggie & -0.364 & $2.46
      \times 10^{-7}$ & N/A\\ organic & -0.361 & $3.13 \times 10^{-7}$
      & 6.32\\ booze & -0.360 & $3.34 \times 10^{-7}$ & N/A\\ grill &
      -0.354 & $5.4 \times 10^{-7}$ & 6.24\\ chocolate & -0.351 &
      $6.77 \times 10^{-7}$ & 7.86\\ \#vegan & -0.350 & $7.47 \times
      10^{-7}$ & N/A\\ \hline \hline mcdonalds & 0.246 & $6.18 \times
      10^{-4}$ & 5.98\\ eat & 0.241 & $8.22 \times 10^{-4}$ & 7.04\\
      wings & 0.222 & $2.13 \times 10^{-3}$ & 6.52\\ hungry & 0.210 &
      $3.65 \times 10^{-3}$ & 3.38\\ heartburn & 0.194 & $7.37 \times
      10^{-3}$ & N/A\\ ham & 0.177 & $1.45 \times 10^{-2}$ & 5.66\\
      \hline
    \end{tabular}
  \end{center}
  \caption{The top 25 food-related words only with strongest negative correlation to obesity level (top), and the 6 food-related words with positive correlation to obesity level and $p$-value less than 0.05 (bottom).
    \label{tbl:obesitynegpos}
}
\end{table}

The above analysis demonstrates that different cities have unique
characteristics.
We now ask whether cities can be sorted into groups based
solely upon similarities in their word distributions.
Bettencourt \emph{et al.}~\cite{Bettencourt2010} used data on the
economy, crime and innovation to characterize cities; here we use a
similar methodology except with word frequency data to uncover
so-called `kindred' cities.

\begin{figure*}[tbp!]
  \begin{center}
    \includegraphics[width=2\columnwidth]{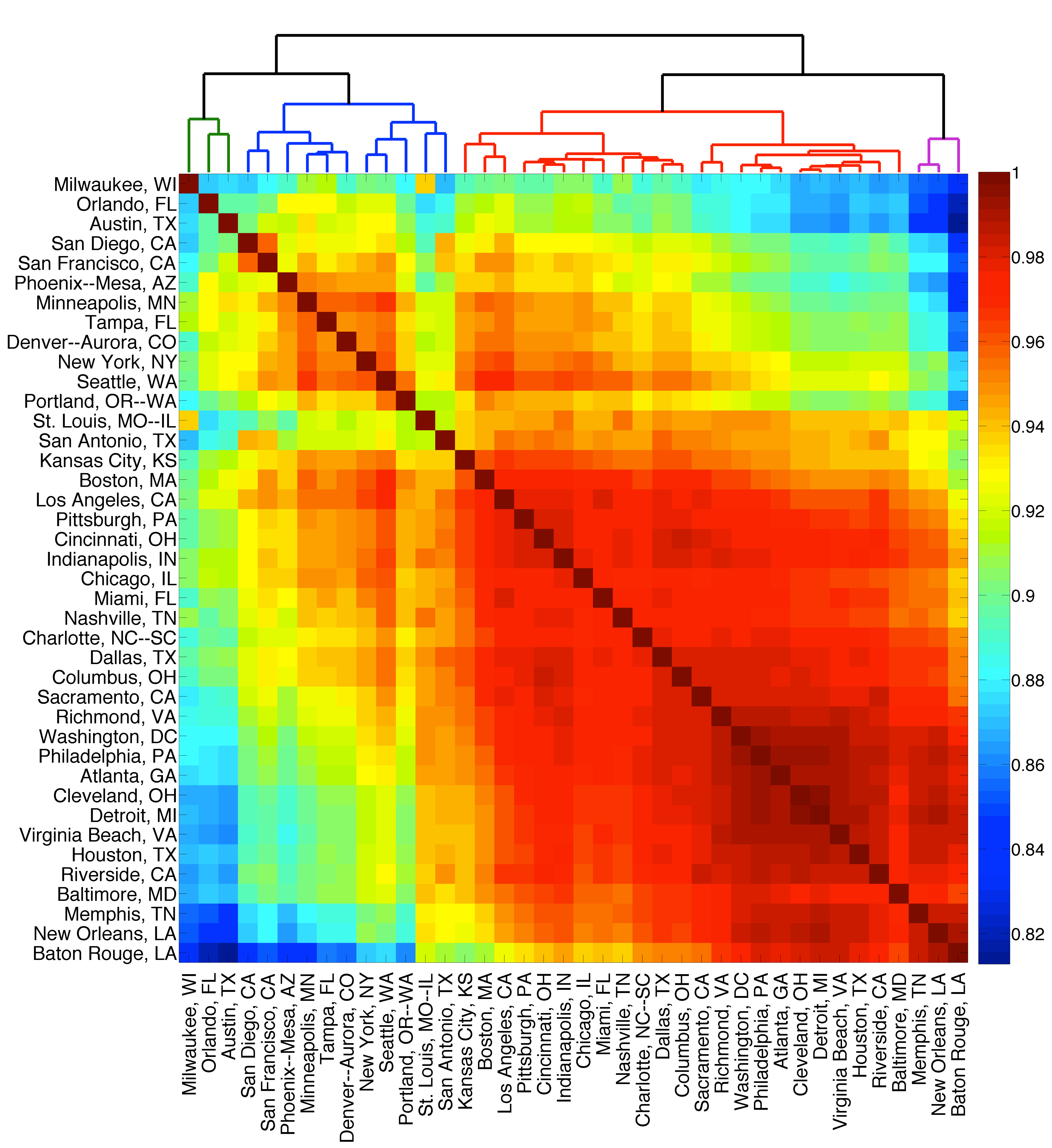}
    \caption{
      Cross-correlations between word frequency distribution
      differences for the 40 cities with highest word counts.
Red
      signifies cities with similar word frequency distribution, while
      blue signifies cities with dissimilar word frequency
      distributions. 
          \label{fig:city_groups}
 }
  \end{center}
\end{figure*}

We group the top 40 cities with highest total word counts in 2011 by
calculating the linear correlation between word frequency vectors
$\mathbf{f}$ as we did in Figure \ref{fig:statesclustergram}.
The resulting cross-correlation
matrix is shown in Figure \ref{fig:city_groups}, with red signifying
strong correlation between cities.
Firstly we note that all cities
show similar word frequency distributions, with all correlations being
higher than $r = 0.8$.
As was the case for the states (see Figure
\ref{fig:statesclustergram}), we see one clear large group of strongly correlated cities emerge
in the lower right corner, with a smaller distinct cluster appearing
at the top left.
Perhaps uniquely, these groupings are defined solely
by similarities in word usage between cities, rather than by geography
or economic indicators.

We cluster cities using an agglomerative hierarchical method with
average linkage clustering~\cite{Jain1999}, as shown in the dendrogram at the top of
Figure \ref{fig:city_groups}, and highlight the 4 clusters with lowest
linkage threshold using different colors.
As one might expect, some
cities that are geographically nearby are grouped together.
Notable
examples are the Southern cities of Baton Rouge, New
Orleans and Memphis in the lower right of the plot, as well as the
Californian cities of San Diego and San Francisco at top left.
However, this pattern does not hold for all cities; while there is the
suggestion of a north/south grouping between the two clusters at the
top left and the two at the bottom right, some cities such as Austin
and Tampa in the south and Detroit and Philadelphia in the north go
against this trend.
The cities of Cleveland and Detroit are the most
alike in word use, having a cross-correlation of $r = 0.995$, while
Austin and Baton Rouge are the most dissimilar with a
cross-correlation of $r = 0.813$.
Indianapolis is the city with
highest average correlation to the word use in other cities
($\bar{r}=0.961$), while Minneapolis shows the most unique word use
on average, with $\bar{r} = 0.884$.

\section{Discussion}    
\label{sec:geotweets.discussion}

In this paper we have examined word use in urban areas in the United
States, using a simple mathematical method which has been shown to
have great flexibility, sensitivity and robustness.
We have used this
tool to map areas of high and low happiness and score individual
states and cities for average word happiness.
In order to understand in greater
detail how word usage influences happiness, we used word shift
graphs to find the words which produced the greatest difference between
the happiness scores of each individual city and the average for the entire US,
and socioeconomic census data to attempt to explain the usage of
certain words.
A significant driver of the happiness score for
individual cities was found to be frequency of profanity; we
believe that future studies of regional variation in swear word use or
`geoprofanity' could help explain geographical differences in
happiness.
Indeed, swearing has previously been found to be a
predictor of large-scale protests and social uprisings in Iran
\cite{Elson2012}.

Happiness within the US was found to correlate strongly with wealth,
showing large positive correlation with increasing household income and
strong negative correlation with increasing poverty.
This is consistent with the first part of the `Easterlin
paradox'~\cite{Easterlin1974}, that within countries at a given time
happiness consistently increases with income.
The second part of the
paradox is that while personal wealth has been observed to
consistently increase over time, happiness has tended to decrease in
both developed and developing countries
\cite{Easterlin1974,Easterlin2010}.
A previous result using this
method showing a decline in happiness over the 2009-2011 period (see
Figure 3 of~\cite{Dodds2011}) is consistent with this finding.
The
relationship between wealth and happiness is still highly debated;
recent works by Stevenson and Wolfers~\cite{Stevenson2008} claim to
show a direct correlation between gross domestic product and
subjective well-being across countries, while Di Tella and MacCulloch
\cite{DiTella2008} in the same year argue that the Easterlin paradox
is in fact exacerbated if other economic variables than just income
are considered.

Interestingly, happiness was also observed to anticorrelate
significantly with obesity.
A similar link between obesity and
happiness has previously been reported~\cite{Fontaine1996},
particularly for individuals who report low self control
\cite{Stutzer2007}.
However, as some authors point out, the presence
of chronic illnesses accompanying obesity can confound the link
between obesity and psychological well-being~\cite{Doll2000}, and
indeed an inverse relationship between weight and depression has been
found in some studies~\cite{Palinkas1996}.
We remark that it should
be possible to use techniques such as those described here to mine
social network data for real-time surveying.
For example, the
potential for identifying areas with high obesity based solely on word
use is significant.

There are a number of legitimate concerns to be raised about how well
the Twitter data set can be said to represent the happiness of the
greater population.
Roughly 15\% of online adults regularly use Twitter,
and 18-29 year-olds and minorities tend to be more highly represented
on Twitter than in the general population~\cite{Smith2012}.
Furthermore, the fact that we collected only around 10\% of all tweets
during the calendar year 2011 means that our data set is a non-uniform
subsample of statements made by a non-representative portion of the
population.

In this work we have only scratched the surface of what is possible
using this particular dataset.
In particular, we have not examined
whether or not these methods have any predictive power---future
research could look at how observed changes in the Twitter data set,
as measured using the hedonometer algorithm, predict changes in the
underlying social and economic characteristics measured using
traditional census methods.  
In particular, we plan to revisit this
study when census data for 2012 becomes available to investigate how
changes in demographics across urban areas is reflected in happiness
as measured by word use.

\acknowledgments
The authors are grateful for the computational resources provided by the Vermont Advanced Computing Core which is supported by NASA (NNX 08A096G), and the Vermont Complex Systems Center. 
LM and CMD were supported by NSF grant DMS-0940271 and PSD was supported by NSF CAREER Award \#0846668. 
The authors also wish to acknowledge support from the MITRE Corporation.

\bibliographystyle{plain}
\bibliography{library,websites}

\clearpage


\appendix

\section*{Appendices}
\section{Data set and states}

\setcounter{figure}{0} \renewcommand{\thefigure}{A\arabic{figure}}
\setcounter{table}{0} \renewcommand{\thetable}{A\arabic{table}}

\begin{figure}[hpb!]
  \begin{center}
    \includegraphics[width=\columnwidth]{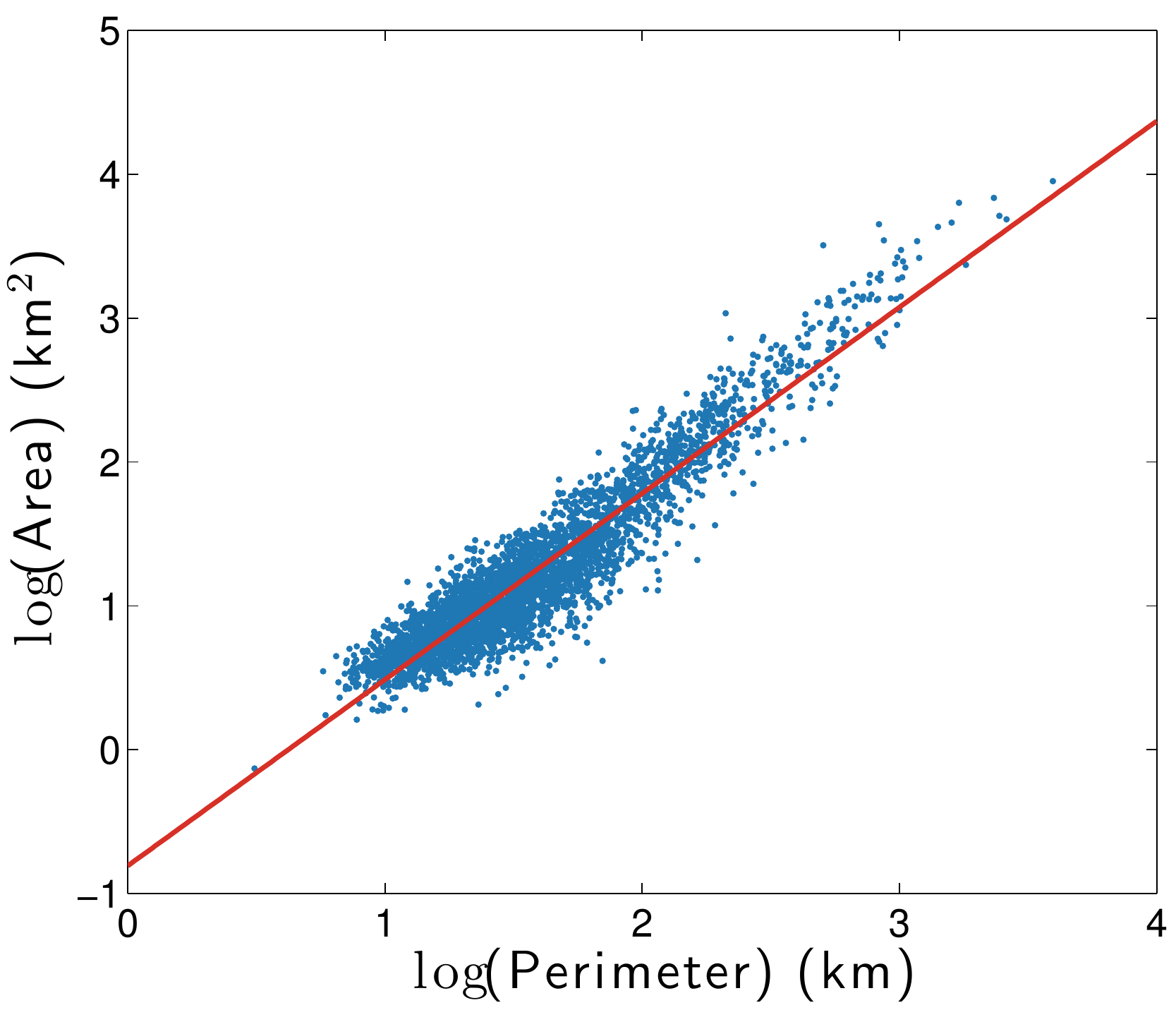}
    \caption{Approximate power law relationship between city area and perimeter for all 3592 cities in the census data set.
The fractal dimension is approximately 1.294, the slope of the red trend line.
    \label{fig:cities_fracdim}
}
  \end{center}
\end{figure}

In figure \ref{fig:cities_fracdim} we show the relationship between
perimeter and area for the 3592 cities in the MAF/TIGER data database,
which follow an approximate power law.  
The
smallest city in both area and perimeter is Richmond, California,
while the largest city is New York, whose perimeter extends far north
into Connecticut and is agglomerated with Newark, New Jersey in this
data set.  
We find that city area shows an approximate power-law
dependence upon perimeter, with an average fractal dimension of
$\alpha = 1.294$.  
Similar results have been reported
previously for cities~\cite{White1993,Shen2002}, and have even been
found to compare well with the fractal dimension of malignant skin
lesions~\cite{Hern2008}.

In preprocessing the Twitter data set we have attempted to remove
tweets from users that are clearly automated bots, in particular
tweets from weather-recording services which periodically report
values of temperature, humidity and the like.
Users for whom more
than 15$\%$ of their tweets contained the words `humid', `humidity',
`pressure' or `earthquake' were removed from the dataset.
The happiness of individual cities tended to be biased towards the score 
for each city name
(as the name of each city was more likely to be found within that city);
to reduce this bias we removed the words `atlantic',
`grand',
`green',
`falls',
`lake',
`new',
`santa',
`haven',
and `battle' from the cities data set.
We also
made the decision to remove all variants of the racial pejorative or
`N-word' from calculations of $\havgfn$.
Variants of this word have
very low happiness values, averaging $\havgfn = 2.92$, and consequently
were found to be highly influential in determining the average city
happiness.
However, when examining individual tweets we found that
this word appeared to be being used in conversation as a more
colloquial stand in for the word `friend' in the vast majority of
cases, and not in fact in any particularly negative sense.  
As such,
we decided that scoring of the word was unfairly biasing our results
towards the negative and removed it.  
Future work will
investigate the scoring of phrases instead of words, which will reduce
the need for this type of adjustment.

\begin{figure}[hbp!]
  \begin{center}
    \includegraphics[width=\columnwidth]{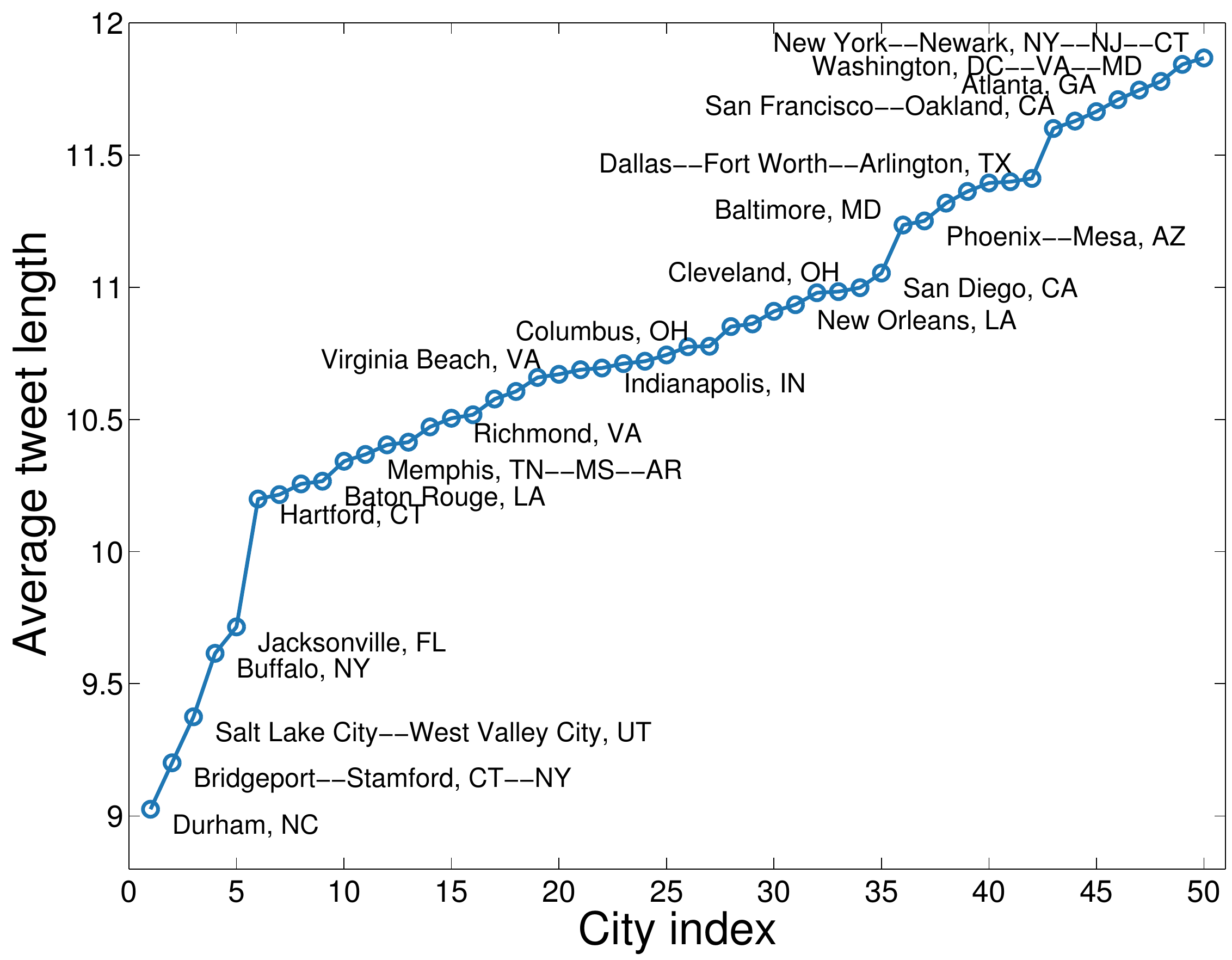}
    \caption{
      Average message length for US cities with more than 50000 LabMT
      words collected during 2011.
          \label{fig:tweetlength}
}
  \end{center}
\end{figure}

For each city we create the normalized word frequency distribution
$\hat{f}(i) = f_i / n$, where $n$ is the total number of tweets
collected for that city.
The sum $\sum_i^N f_i /n$ therefore
represents the average number of LabMT words per tweet, the mean of
which is approximately 7.1.
In figure \ref{fig:tweetlength} we show
the average tweet length for the US cities for which we have collected
more than 50000 words throughout 2011.
Average tweet lengths range
from 9 words per tweet for Durham, North Carolina 
up to almost 12 words per tweet in
New York.

\begin{figure*}[tbp!]
  \begin{center}
    \includegraphics[width=2\columnwidth]{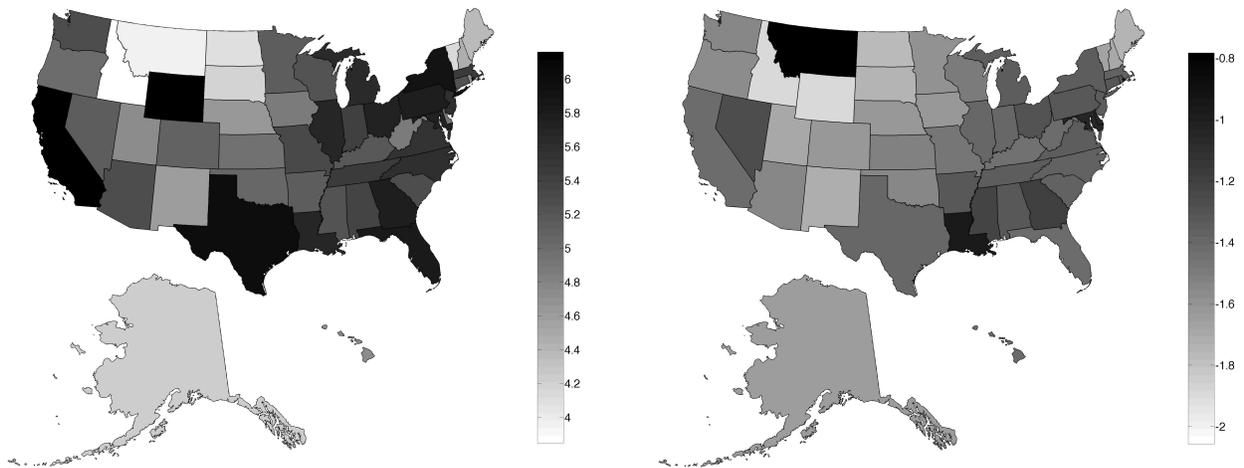}
    \caption{
      Choropleths showing (the base-10 logarithm of) raw count (left)
      and number of geotagged tweets collected per capita (right) in
      each US state during the calendar year 2011.
          \label{fig:states_nwords}
    }
  \end{center}
\end{figure*}

Figure \ref{fig:states_nwords} shows choropleths for the number of
geotagged tweets collected (left) and number of geotagged tweets
normalized by state population (right) for the 2011 data set.
In both
plots the gray scale is logarithmic.
In table \ref{tbl:hap_states}
we show the complete list of happiness scores for all US states.
Word
shift plots for each state are presented in Appendix B (online)~\cite{geotweets-onlinematerial}.

In tables \ref{tbl:obesityposcorr} and \ref{tbl:obesitynegcorr} we
show lists of the top 25 LabMT words with highest positive and
negative correlation to obesity, respectively.
In table
\ref{tbl:obesitynocorr} we show the words with lowest correlation to
obesity, that is, the words with $p$-values greater than 0.9.
Complete lists for for word correlations with all demographic
attributes can be found in Appendix D (online)~\cite{geotweets-onlinematerial}.

\section*{B,C,D,E,F\quad Online appendices}

The remaining appendices are located online, at
\url{http://www.uvm.edu/storylab/share/papers/mitchell2013a/}. 
Appendix B contains word shift graphs for all states, 
Appendix C contains a comparison between happiness and the Gallup-Healthways well-being
measure as well as tweet maps and word shift graphs for all cities, 
and
Appendix D contains complete tables of correlations between
demographic attributes and both happiness and word usage.
Appendix E contains the complete list of LabMT words ordered by correlation with happiness, and
Appendix F is a daily-updating happiness map of the United States.

\clearpage

\begin{table}[htb]
  \begin{center}
    \begin{tabular}{|c|l|c|}
      \hline
      Rank & State    & $\havgfn$\\
      \hline
      1 & Hawaii & 6.17\\
      2 & Maine & 6.14\\
      3 & Nevada & 6.12\\
      4 & Utah & 6.11\\
      5 & Vermont & 6.11\\
      6 & Colorado & 6.10\\
      7 & Idaho & 6.10\\
      8 & New Hampshire & 6.09\\
      9 & Washington & 6.08\\
      10 & Wyoming & 6.08\\
      11 & Minnesota & 6.07\\
      12 & Arizona & 6.07\\
      13 & California & 6.07\\
      14 & Florida & 6.06\\
      15 & New York & 6.06\\
      16 & New Mexico & 6.05\\
      17 & Iowa & 6.05\\
      18 & Oregon & 6.05\\
      19 & North Dakota & 6.04\\
      20 & Nebraska & 6.04\\
      21 & Wisconsin & 6.03\\
      22 & Kansas & 6.03\\
      23 & Alaska & 6.02\\
      24 & Oklahoma & 6.02\\
      25 & Massachusetts & 6.02\\
      26 & Montana & 6.01\\
      27 & Missouri & 6.01\\
      28 & Kentucky & 6.00\\
      29 & New Jersey & 5.99\\
      30 & West Virginia & 5.99\\
      31 & Illinois & 5.99\\
      32 & Rhode Island & 5.99\\
      33 & Indiana & 5.98\\
      34 & Texas & 5.98\\
      35 & South Dakota & 5.98\\
      36 & Virginia & 5.97\\
      37 & Tennessee & 5.97\\
      38 & Connecticut & 5.97\\
      39 & Pennsylvania & 5.97\\
      40 & South Carolina & 5.96\\
      41 & North Carolina & 5.96\\
      42 & Ohio & 5.96\\
      43 & Arkansas & 5.95\\
      44 & District of Columbia & 5.94\\
      45 & Michigan & 5.94\\
      46 & Alabama & 5.94\\
      47 & Georgia & 5.94\\
      48 & Delaware & 5.92\\
      49 & Maryland & 5.90\\
      50 & Mississippi & 5.89\\
      51 & Louisiana & 5.88\\

      \hline
    \end{tabular}
  \end{center}
  \caption{Happiness scores $\havgfn$ for each US state, in order from highest to lowest.
    \label{tbl:hap_states}
}
\end{table}

\begin{table}
  \begin{center}
    \begin{tabular}{|l|c|c|c|}
      \hline Word & $\rho$ & $p$-value & $\havgfn(w_i)$\\ \hline don't &
      0.461 & $2.28 \times 10^{-11}$ & 3.70\\ give & 0.443 & $1.57
      \times 10^{-10}$ & 6.54\\ lie & 0.442 & $1.68 \times 10^{-10}$ &
      2.60\\ hell & 0.438 & $2.56 \times 10^{-10}$ & 2.22\\ my & 0.438
      & $2.74 \times 10^{-10}$ & 6.16\\ she & 0.433 & $4.36 \times
      10^{-10}$ & 6.18\\ okay & 0.423 & $1.18 \times 10^{-9}$ & 6.56\\
      like & 0.419 & $1.72 \times 10^{-9}$ & 7.22\\ girl & 0.419 &
      $1.76 \times 10^{-9}$ & 7.00\\ know & 0.415 & $2.54 \times
      10^{-9}$ & 6.10\\ act & 0.412 & $3.48 \times 10^{-9}$ & 6.00\\
      bitch & 0.411 & $4.01 \times 10^{-9}$ & 3.14\\ me & 0.403 & $8.5
      \times 10^{-9}$ & 6.58\\ all & 0.400 & $1.08 \times 10^{-8}$ &
      6.22\\ nothin & 0.399 & $1.14 \times 10^{-8}$ & 3.64\\ better &
      0.398 & $1.34 \times 10^{-8}$ & 7.00\\ bored & 0.396 & $1.5
      \times 10^{-8}$ & 3.04\\ bed & 0.395 & $1.72 \times 10^{-8}$ &
      7.18\\ sleep & 0.395 & $1.78 \times 10^{-8}$ & 7.16\\ wish &
      0.388 & $3.25 \times 10^{-8}$ & 6.92\\ never & 0.387 & $3.43
      \times 10^{-8}$ & 3.34\\ money & 0.380 & $6.41 \times 10^{-8}$ &
      7.30\\ hate & 0.378 & $7.57 \times 10^{-8}$ & 2.34\\ make &
      0.376 & $9.32 \times 10^{-8}$ & 6.00\\ cant & 0.376 & $9.33
      \times 10^{-8}$ & 3.48\\ \hline
    \end{tabular}
  \end{center}
  \caption{
    Top 25 words with strongest positive Spearman correlation $\rho$
    to obesity in 2011.
Stop words with $4 < \havgfn < 6$ have been removed
    from the list.  
    \label{tbl:obesityposcorr}
  }
\end{table}

\begin{table}
  \begin{center}
    \begin{tabular}{|l|c|c|c|}
      \hline
      Word    & $\rho$        & $p$-value & $\havgfn(w_i)$\\
      \hline
      cafe & -0.509 & $6.07 \times 10^{-14}$ & 6.78\\
      photo & -0.493 & $4.87 \times 10^{-13}$ & 6.88\\
      thai & -0.476 & $3.69 \times 10^{-12}$ & 6.22\\
      fitness & -0.472 & $5.92 \times 10^{-12}$ & 6.92\\
      park & -0.468 & $9.59 \times 10^{-12}$ & 7.08\\
      yoga & -0.448 & $8.82 \times 10^{-11}$ & 7.04\\
      restaurant & -0.448 & $8.93 \times 10^{-11}$ & 7.06\\
      banana & -0.434 & $3.77 \times 10^{-10}$ & 6.86\\
      event & -0.433 & $4.54 \times 10^{-10}$ & 6.12\\
      hotel & -0.429 & $6.41 \times 10^{-10}$ & 6.16\\
      spa & -0.420 & $1.54 \times 10^{-9}$ & 6.92\\
      interesting & -0.420 & $1.62 \times 10^{-9}$ & 7.52\\
      design & -0.409 & $4.76 \times 10^{-9}$ & 6.32\\
      apple & -0.408 & $5.22 \times 10^{-9}$ & 7.44\\
      feliz & -0.406 & $6.47 \times 10^{-9}$ & 6.04\\
      photos & -0.404 & $7.8 \times 10^{-9}$ & 6.94\\
      wine & -0.400 & $1.08 \times 10^{-8}$ & 6.42\\
      bike & -0.399 & $1.22 \times 10^{-8}$ & 6.72\\
      sun & -0.398 & $1.31 \times 10^{-8}$ & 7.80\\
      delicious & -0.392 & $2.17 \times 10^{-8}$ & 7.92\\
      flight & -0.391 & $2.34 \times 10^{-8}$ & 6.06\\
      sunset & -0.391 & $2.51 \times 10^{-8}$ & 7.16\\
      lounge & -0.389 & $2.93 \times 10^{-8}$ & 6.50\\
      mortgage & -0.386 & $3.83 \times 10^{-8}$ & 3.88\\
      dinner & -0.386 & $3.85 \times 10^{-8}$ & 7.40\\
      \hline
    \end{tabular}
  \end{center}
  \caption{
    Top 25 words with strongest negative Spearman correlation $\rho$
    to obesity in 2011.
    Stop words with $4 < \havgfn < 6$ have been
    removed from the list.
    \label{tbl:obesitynegcorr}
}
\end{table}

\begin{table}
  \begin{center}
    \begin{tabular}{|l|c|c|c|}
      \hline
      Word    & $\rho$        & $p$-value & $\havgfn(w_i)$\\
      \hline
      olive & -0.001 & $9.94 \times 10^{-1}$ & 6.00\\
      refrigerator & 0.001 & $9.9 \times 10^{-1}$ & N/A\\
      hashbrowns & 0.002 & $9.83 \times 10^{-1}$ & N/A\\
      eatting & -0.002 & $9.76 \times 10^{-1}$ & N/A\\
      sauteed & 0.003 & $9.72 \times 10^{-1}$ & N/A\\
      fritos & -0.003 & $9.69 \times 10^{-1}$ & N/A\\
      munch & 0.003 & $9.64 \times 10^{-1}$ & N/A\\
      doughnuts & -0.003 & $9.62 \times 10^{-1}$ & N/A\\
      cola & -0.004 & $9.62 \times 10^{-1}$ & N/A\\
      okra & -0.004 & $9.59 \times 10^{-1}$ & N/A\\
      grapes & 0.004 & $9.51 \times 10^{-1}$ & N/A\\
      noodles & -0.004 & $9.51 \times 10^{-1}$ & N/A\\
      quiznos & 0.005 & $9.49 \times 10^{-1}$ & N/A\\
      cucumbers & 0.005 & $9.46 \times 10^{-1}$ & N/A\\
      chow & 0.006 & $9.3 \times 10^{-1}$ & N/A\\
      walnut & 0.007 & $9.28 \times 10^{-1}$ & N/A\\
      mulberry & 0.007 & $9.19 \times 10^{-1}$ & N/A\\
      muesli & 0.008 & $9.17 \times 10^{-1}$ & N/A\\
      hershey's & 0.008 & $9.17 \times 10^{-1}$ & N/A\\
      snickers & 0.008 & $9.16 \times 10^{-1}$ & N/A\\
      krispy & -0.008 & $9.15 \times 10^{-1}$ & N/A\\
      nugget & -0.008 & $9.12 \times 10^{-1}$ & N/A\\
      smores & 0.008 & $9.1 \times 10^{-1}$ & N/A\\
      popcorn & 0.009 & $9.07 \times 10^{-1}$ & 6.76\\
      \hline
    \end{tabular}
  \end{center}
  \caption{
    The 24 food-related words which show least correlation with
    obesity, and have $p$-values greater than 0.9.  
    Words are arranged
    in decreasing order of $p$-value.
    \label{tbl:obesitynocorr}
}
\end{table}

\clearpage

\end{document}